\documentclass[twocolumn,times]{aastex7}
\usepackage{times}
\graphicspath{{./}}

\usepackage{amsmath}
\allowdisplaybreaks[4]
\usepackage{mathtools}
\usepackage{bm}
\usepackage{subcaption}
\usepackage{mathrsfs}
\usepackage{CJK}

\usepackage{listings}

\begin{document}
\begin{CJK*}{UTF8}{gbsn}

\title{Characterization of Numerical Dissipation in Simulations of Magnetohydrodynamic Turbulence}

\author[orcid=0000-0002-3365-5279,gname=Yuyang, sname=Hua]{Yuyang Hua (华宇阳)}
\affiliation{Center for Applied Physics and Technology, HEDPS and SKLNPT, School of Physics, Peking University, Beijing 100871, China}
\email[show]{yuyanghua@stu.pku.edu.cn}

\author[orcid=0000-0003-2051-1740,gname=Zhonghai, sname=Zhao]{Zhonghai Zhao (赵忠海)}
\affiliation{Center for Applied Physics and Technology, HEDPS and SKLNPT, School of Physics, Peking University, Beijing 100871, China}
\email{zhaozhonghi@pku.edu.cn}

\author[orcid=0000-0001-9186-8574,gname=Bin,sname=Qiao]{Bin Qiao (乔宾)}
\affiliation{Center for Applied Physics and Technology, HEDPS and SKLNPT, School of Physics, Peking University, Beijing 100871, China}
\affiliation{Frontiers Science Center for Nano-optoelectronics, Peking University, Beijing 100094, China}
\email[show]{bqiao@pku.edu.cn}

\begin{abstract}
    Comprehensive characterization of numerical dissipation is essential for high-fidelity simulations of magnetohydrodynamic (MHD) turbulence. In this work, we present an \textit{a posteriori} framework for directly estimating numerical dissipation in MHD turbulence from simulation data without invoking \textit{a priori} assumptions. Implemented in the open-source Python package \texttt{PyMHD}, the framework is applied to simulations of Alfv\'enic turbulence, turbulent small-scale dynamos, and MRI-driven turbulence, yielding a systematic characterization of the anisotropy and spectral properties of numerical dissipation across these regimes. The results indicate that numerical dissipation primarily dissipates energy transferred by the turbulent cascade at small scales, consistent with the conventional interpretation. However, its spectral properties are distinct from those of physical viscosity and resistivity, such that it cannot simply be represented by effective dissipation coefficients. In addition, numerical dissipation inherits the anisotropy of the underlying turbulence, and can even exhibit anomalous anti-dissipative behavior under certain circumstances. Moreover, this framework enables identification of the conditions under which physical dissipation dominates numerical dissipation across all scales, thereby providing practical guidance for achieving high-fidelity simulations of astrophysical MHD turbulence.
\end{abstract}
\keywords{
\uat{Plasma astrophysics}{1261} ---
\uat{Astrophysical fluid dynamics}{101} ---
\uat{Magnetohydrodynamics}{1964} ---
\uat{Computational methods}{1965} ---
\uat{Astronomy software}{1855}
}


\section{Introduction}\label{sec:introduction}

Magnetohydrodynamic (MHD) turbulence is ubiquitous in astrophysics \citep{Biskamp2003, Zhou2004, Schekochihin2022}, and a comprehensive understanding of its fundamental properties is crucial for a broad range of astrophysical applications. Prominent examples include turbulent interstellar and intergalactic media (ISM/IGM) \citep{Elmegreen2004, Hu2019, Beattie2025}, turbulent angular momentum transport driven by magnetorotational instability (MRI) \citep{Balbus1991, Balbus1998, Kawazura2024}, magnetic-field amplification via turbulent dynamos \citep{Schekochihin2004, Tobias2021, Warnecke2023, Zhao2024}, and turbulent magnetic reconnection in solar flares \citep{Somov2013, Ping2023, Wang2023}.

Despite the emergence of laboratory plasma experiments \citep{Tzeferacos2018, Bott2021, Bott2022}, numerical simulation remains indispensable for MHD turbulence research. Direct numerical simulation (DNS), in particular, provides the most detailed physics by resolving the entire turbulent cascade \citep{Moin1998, Dong2022}. However, astrophysical plasmas are characterized by high Reynolds numbers. For example, the Reynolds number in the ISM ranges from $10^6$ to $10^9$ \citep{Beattie2025}, while the magnetic Reynolds number in the MRI-unstable, neutrino-cooled interiors of protoneutron stars can reach up to $\mathrm{Rm} = 10^{19}$ \citep{Held2022}. According to classical turbulence phenomenology \citep[e.g.,][]{Frisch1995}, such high Reynolds numbers imply a vast range of scales, which often renders DNS computationally intractable. To address this, large-eddy simulations (LES) \citep{Sagaut2006, Miesch2015} have been introduced to solve filtered equations on coarser grids, while modeling unresolved subgrid-scale (SGS) dynamics via closure models.

High-fidelity DNS and LES rely heavily on the characterization of errors introduced by numerical schemes. Specifically, numerical errors in grid-based codes are twofold \citep{Jasak1996}: domain discretization errors originating from finite grid spacing and discrete time steps, and truncation errors introduced by approximating differential operators in the governing equations using finite-order interpolation or reconstruction. While domain discretization errors are well accounted for in relevant studies \citep{Moin1998, Guilet2022}, the impact of truncation errors is often less scrutinized. However, these errors can significantly contaminate the small-scale dynamics of turbulence, especially in LES \citep{Ghosal1996, Chow2003, Dairay2017, Komen2017}.

Modified equation analysis (MEA) is a classical technique for analyzing truncation errors, in which they take the form of truncation terms that augment the original PDEs, yielding the modified equations \citep{Hirt1968, Griffiths1986, Grinstein2007}. Aside from round-off error, the modified equations represent the approximate PDEs that the numerical solution actually satisfies \citep{Warming1974}. Furthermore, \citet{Grinstein2007} demonstrated that all conservative numerical schemes have truncation terms in the form of the divergence of stress tensors. For certain schemes, these stress tensors closely resemble the SGS stress in the context of LES. This led to the idea of implicit LES (ILES), in which the governing PDEs are solved on coarser grids, while the SGS dynamics are captured by truncation terms without incorporating explicit models \citep{Boris1992, Grinstein2007}. Alternatively, stability constraints require truncation terms to be intrinsically dissipative, leading to the concept of numerical dissipation. These two seemingly distinct interpretations of truncation terms are consistent in the sense that both mechanisms act to dissipate turbulent energy at small scales. Indeed, some studies adopt these concepts interchangeably, treating ideal simulations characterized by effective Reynolds numbers as ILES \citep{Beattie2025, Shivakumar2025}.

In principle, MEA offers an analytical framework for characterizing numerical dissipation, in which the truncation terms can be explicitly derived via Taylor expansion \citep{Warming1974}. However, this analytical approach becomes mathematically infeasible when applied to multidimensional, highly nonlinear systems \citep{Schranner2015}, including the MHD turbulence considered in this work. Consequently, this limitation of MEA motivates the adoption of alternative techniques capable of estimating numerical dissipation \textit{a posteriori} in complex turbulent flows. These techniques generally fall into three main categories: empirical ansatz methods, spectral energy budget methods, and residual-based methods in physical space.

The empirical ansatz methods are based on \textit{a priori} assumptions regarding the numerical dissipation or the underlying MHD turbulence. \citet{Rembiasz2017} proposed an empirical ansatz where the numerical dissipation coefficients are expressed as functions of the grid spacing and time step with free parameters. These parameters are subsequently determined by comparing the numerical simulation results with analytical solutions in certain resistive-viscous MHD flows. More recently, \citet{Shivakumar2025} estimated numerical viscosity and resistivity across subsonic and supersonic regimes by fitting analytical models to the turbulent energy spectra of ideal MHD turbulence simulations. Using the empirical relations between dissipation scales and (magnetic) Reynolds numbers established by \citet{Kriel2022, Kriel2025}, they determined the effective (magnetic) Reynolds numbers associated with numerical dissipation. However, both methods rely on the assumption that the behavior of truncation terms is indeed analogous to the physical dissipation, and the robustness of their underlying empirical models across diverse flow regimes remains uncertain. Consequently, these assumptions may introduce considerable uncertainties, as reflected in, for instance, the estimated value of the magnetic Prandtl number $\mathrm{Pm} = 6.2^{+5.6}_{-4.8}$ in the supersonic regime as reported in \citet{Shivakumar2025}.

Spectral energy budget methods attribute the residuals of the spectral energy transfer equations in ILES to numerical dissipation \citep{Domaradzki2003, Zhou2014, Grete2023}. The numerical dissipation coefficients are subsequently determined by comparing the residuals to the contribution of physical dissipation terms within the energy transfer equations. While \citet{Domaradzki2003} and \citet{Zhou2014} focus mainly on hydrodynamic turbulence, \citet{Grete2023} demonstrated in simulations of MHD turbulence that the effective viscosity and resistivity coefficients for numerical dissipation scale with grid spacing $\Delta$ as $\nu^{\mathrm{eff}}\propto \Delta^{1.22}$ and $\eta^{\mathrm{eff}}\propto \Delta^{1.34}$, respectively. However, this methodology is inherently restricted to homogeneous, isotropic turbulence with periodic boundary conditions, as it relies on the shell-to-shell energy transfer analysis in Fourier space \citep{Grete2017}. Consequently, this framework is not directly applicable to complex geometries, strongly anisotropic flows, or MHD turbulence exhibiting inverse energy cascades \citep{Grete2023}.

The residual-based methods in physical space were originally proposed by \citet{Schranner2015}, where the residual of the kinetic energy transport equation is estimated locally in physical space for arbitrary subdomains down to the single-cell level. This framework was subsequently applied to realistic flow configurations targeting industrial applications \citep{Castiglioni2015, Cadieux2017, Komen2017}. Furthermore, \citet{Castiglioni2019} generalized this framework to arbitrary PDEs and extended the analysis to include dispersive errors. However, traditional residual-based methods rely on the introduction of a (pseudo-)kinetic energy and a corresponding scalar energy transport equation. This scalar formulation limits its capability to capture the anisotropic characteristics of numerical dissipation, which are crucial in anisotropic MHD turbulence. Additionally, the accuracy of these residual-based methods is contingent upon the specific implementation of high-order schemes employed to evaluate the residual terms. Specifically, \citet{Castiglioni2019} utilized a spectral collocation method for periodic boundary conditions, while resorting to the sixth-order compact finite difference scheme for non-periodic domains. Due to the absence of inherent shock-capturing capabilities, both schemes are prone to inducing spurious oscillations in the presence of discontinuities. As a result, applying them directly to compressible turbulence containing shocks may compromise the accuracy of the numerical dissipation estimation.

In this paper, we develop a novel \textit{a posteriori} framework for the direct estimation of numerical dissipation, with a specific focus on simulations of MHD turbulence, including Alfv\'enic MHD turbulence, turbulent small-scale dynamos (SSDs), and MRI-driven turbulence. The objectives of this work are twofold: first, to establish a residual-based method for estimating numerical dissipation without invoking any \textit{a priori} empirical ansatz or restrictive assumptions, while enabling the characterization of its anisotropy; second, to apply this framework to simulations of astrophysical MHD turbulence and characterize the properties of numerical dissipation across various regimes. This analysis provides practical guidance for setting simulation parameters in high-fidelity DNS of MHD turbulence.

To this end, building upon the general method of \citet{Castiglioni2019}, we develop a residual-based framework that provides a more comprehensive characterization of numerical dissipation. Instead of estimating numerical dissipation from the residual of a scalar energy transport equation as in \citet{Castiglioni2019}, we directly evaluate the residuals of the governing MHD equations, thereby extracting the numerical viscous term $\bm{D}^{\mathrm{num}}_{\mathrm{vis}}$ from the momentum equation and the numerical resistive term $\bm{D}^{\mathrm{num}}_{\mathrm{res}}$ from the induction equation. This formulation preserves the vector form of the residual terms and thus enables us to quantify the anisotropic characteristics of numerical dissipation. To improve the accuracy and robustness of residual evaluation, spatial derivatives are computed using a novel high-order targeted compact scheme (TCS) with high spectral resolution and robust shock-capturing capability, while temporal derivatives are evaluated using Fornberg's finite difference formulas on arbitrarily spaced grids \citep{Fornberg1988} to accommodate the variable time steps determined by the CFL condition. We further characterize the spectral properties of numerical dissipation by defining the component-wise viscous and resistive dissipation spectra, thereby enabling a systematic analysis of how numerical dissipation shapes the turbulent cascade. We implement this framework in an open-source Python package \texttt{PyMHD}\footnote{\url{https://github.com/PlasmaHua/PyMHD}} \citep{Hua2026}, a general-purpose post-processing toolkit for simulations of MHD turbulence. To the best of our knowledge, this work presents the first residual-based estimation of numerical dissipation for MHD simulations and provides the first open-source implementation for estimating numerical dissipation in turbulence simulations.

The remainder of this paper is organized as follows. Section~\ref{sec:methods} introduces the proposed residual-based framework for numerical dissipation estimation. Section~\ref{sec:properties} applies the framework to simulations of MHD turbulence and analyzes the anisotropy and spectral properties of numerical dissipation. Section~\ref{sec:dns} further applies the proposed framework to provide practical guidance for achieving high-fidelity DNS in simulations of turbulent MRI-driven dynamos. Finally, Section~\ref{sec:conclusions} summarizes the main conclusions and outlines future extensions.

\section{Methods} \label{sec:methods}

\subsection{Simulations of MHD Turbulence}

\subsubsection{Governing Equations}

The macroscopic dynamics of collisional plasmas are governed by the resistive-viscous MHD equations:
\begin{gather}
    \frac{\partial \rho}{\partial t}+\nabla \cdot (\rho \bm{u}) = 0, \\
    \label{eq:momentum} \rho \left(\frac{\partial \bm{u}}{\partial t} +\bm{u}\cdot \nabla \bm{u} \right) =
    - \nabla p + \bm{J} \times \bm{B} + \nu \nabla \cdot \mathbb{T}, \\
    \label{eq:induction} \frac{\partial \bm{B}}{\partial t} = \nabla \times (\bm{u}\times \bm{B}) + \eta \nabla^2\bm{B},\\
    \nabla \cdot \bm{B} = 0,
\end{gather}
where $\rho$, $p$, $\bm{u}$, and $\bm{B}$ denote the density, pressure, velocity, and magnetic field, respectively, and $\bm{J} = \nabla \times \bm{B}$ is the current density. The viscous stress tensor $\mathbb{T}$ is defined as\footnote{In this work, the coefficient $\nu$ is not included in the definition of the viscous stress tensor.}
\begin{equation}
\mathbb{T} = \rho \left[\nabla \bm{u} + (\nabla \bm{u})^{\top} - \frac{2}{3}\left(\nabla \cdot \bm{u}\right)\mathbb{I} \right],
\end{equation}
where $\mathbb{I}$ denotes the identity tensor, while $\nu$ and $\eta$ are the constant kinematic viscosity and magnetic diffusivity (resistivity), respectively.

For an adiabatic equation of state (EoS), the system is closed by the energy conservation equation:
\begin{equation}\label{eq:energy-conservation}
    \begin{aligned}
    \frac{\partial e}{\partial t} + \nabla \cdot (e\bm{u}) = \nabla \cdot \bigg[ & - \left(p + \frac{1}{2}B^2\right) \bm{u} + \bm{B}(\bm{B} \cdot \bm{u}) \\
    & + \nu\mathbb{T} \cdot \bm{u} - \eta \bm{J} \times \bm{B} \bigg],
    \end{aligned}
\end{equation}
where $e = \epsilon + \rho u^2/2 + B^2/2$ is the total energy density, and $\epsilon = p/(\gamma-1)$ is the internal energy density with $\gamma$ the adiabatic index. For an isothermal EoS where the total energy is not conserved, the energy conservation equation~\eqref{eq:energy-conservation} is replaced with $p = c_{\mathrm{s}}^2\rho$, where $c_{\mathrm{s}}$ represents the constant speed of sound.

\subsubsection{Driven MHD Turbulence}
In driven (forced) MHD turbulence, energy is continuously injected at large scales to sustain a statistically stationary turbulent state. Consequently, the momentum conservation equation~\eqref{eq:momentum} is modified into
\begin{equation}\label{eq:momentum-driven}
    \rho \left(\frac{\partial \bm{u}}{\partial t} + \bm{u}\cdot \nabla \bm{u} \right)
    = - \nabla p + \bm{J} \times \bm{B} + \bm{f}_{\mathrm{drive}} + \nu \nabla \cdot \mathbb{T},
\end{equation}
where $\bm{f}_{\mathrm{drive}}$ is a stochastic driving force generated by an Ornstein-Uhlenbeck (OU) process \citep{Uhlenbeck1930, Schmidt2008, Grete2018}. The driving force is characterized by the autocorrelation time $T_{\mathrm{corr}}$ of the OU process, the energy injection rate $\dot{\varepsilon}_{\mathrm{inj}}$ or the root-mean-square acceleration $a_{\mathrm{rms}}$, and the characteristic injection scale $L_{\mathrm{inj}}$.

The dynamics of driven MHD turbulence depend on the mean magnetic field threading the computational domain \citep{Cho2026}: A strong mean field leads to the regime of Alfv\'enic MHD turbulence, in which the turbulent cascade is governed primarily by nonlinear interactions of Alfv\'en wave packets. In contrast, for a weak (or vanishingly small) mean field, the turbulent velocity fluctuations can amplify the initial seed magnetic field via a small-scale dynamo (SSD), eventually reaching a saturated state where magnetic and kinetic energies are in approximate equipartition \citep{Beresnyak2012, Chirakkara2021}.

In this work, we focus on simulations of Alfv\'enic MHD turbulence and turbulent SSDs in an $L^3$ cubic box with periodic boundary conditions. The former provides a canonical example of anisotropic MHD turbulence \citep{Cho2000}, whereas the latter yields an isotropic turbulent state, allowing us to demonstrate the applicability of the proposed framework to simulations of both anisotropic and isotropic turbulence.

\subsubsection{MRI-driven Turbulence}\label{sec:mri-driven-turbulence}

The local shearing-box approximation with orbital advection is widely adopted in simulations of MRI-driven turbulence \citep{Hawley1995, Masset2000}. We employ the standard Cartesian coordinate system where $x,y,z$ represent radial, azimuthal, and vertical directions, respectively. In this framework, the fluid velocity $\bm{u}$ is decomposed into a background shear flow $\bm{u}_0 = -q\Omega_{0} x \bm{e}_y$ and a velocity fluctuation $\bm{v}$ (i.e., $\bm{u} = \bm{u}_0 + \bm{v}$). Here, $\bm{\Omega}_0 = \Omega_{0} \bm{e}_z$ denotes the local angular velocity, and $q = -{{\mathrm{d}\ln \Omega(r)}}/{{\mathrm{d}\ln r}}$ is the shearing parameter with $q=3/2$ representing the Keplerian profile. In this rotating frame, the momentum and induction equations take the form:
\begin{align}
    \rho \left(\frac{\partial \bm{v}}{\partial t} + \bm{v}\cdot \nabla \bm{v} \right)
    = & - \rho(\bm{u}_{0}\cdot \nabla \bm{v} + \bm{v}\cdot \nabla \bm{u}_{0} + 2\bm{\Omega}_{0} \times \bm{v}) \nonumber\\
    & - \nabla p + \bm{J} \times \bm{B} + \nu \nabla \cdot \mathbb{T},
    \label{eq:mri-momentum}\\
    \frac{\partial \bm{B}}{\partial t} = \nabla \times (\bm{v} \times & \bm{B}) + \nabla \times (\bm{u}_0\times \bm{B}) + \eta \nabla^2\bm{B}.
\end{align}

Anisotropy is a distinct feature of MRI-driven turbulence, which arises from the stretching of magnetic field lines by the background shear \citep{Murphy2015}. This anisotropy leads to the generation of anisotropic Reynolds and Maxwell stresses, which drive the outward angular momentum transport that enables accretion onto compact objects. To achieve high-fidelity DNS of such anisotropic turbulence, it is essential to capture the anisotropic characteristics of numerical dissipation.

In this work, we focus on simulations of isothermal turbulent MRI-driven dynamos in the unstratified Keplerian shearing box of size $L_x \times L_y \times L_z = H \times 2H \times 0.5H$ with resolutions of $N_x \times N_y \times N_z = N \times N \times 0.5N$. Here, $H = c_{\mathrm{s}}/\Omega_{0}$ represents the characteristic scale height of an isothermal disk with sound speed $c_{\mathrm{s}}$, and $N$ denotes the resolution per scale height. In local simulations of turbulent MRI-driven dynamos, the zero-net-flux (ZNF) configuration $\bm{B} = B_0\bm{e}_z\sin(2\pi x/L_x)$ is adopted as the initial condition. With the sound speed $c_{\mathrm{s}}$ taken as the characteristic velocity, the Reynolds number, magnetic Reynolds number, and magnetic Prandtl number are defined as
\begin{equation}
\mathrm{Re} = \frac{c_{\mathrm{s}}H}{\nu},\
\mathrm{Rm} = \frac{c_{\mathrm{s}}H}{\eta},\
\mathrm{Pm} = \frac{\mathrm{Rm}}{\mathrm{Re}}=\frac{\nu}{\eta},
\end{equation}
respectively.

\subsubsection{Godunov MHD Codes}

In this work, we characterize the numerical dissipation introduced by Godunov-type methods in simulations of astrophysical MHD turbulence discussed above. Godunov codes have become the \textit{de facto} standard for astrophysical MHD simulations, with notable examples including \texttt{Athena++} \citep{Stone2020}, \texttt{AthenaK} \citep{Stone2026}, \texttt{AthenaPK} \citep{Grete2021, Holmen2024}, \texttt{PLUTO} \citep{Mignone2007}, \texttt{FLASH} \citep{Fryxell2000}, and more recently, \texttt{Kratos} \citep{Wang2025}. A comparative study by \citet{Joseph2023} has demonstrated that \texttt{Athena++} and \texttt{PLUTO} exhibit similar behavior in terms of numerical viscosity. Therefore, we restrict our analysis to the performance-portable codes \texttt{AthenaK} and \texttt{AthenaPK}, both of which support high-order spatial reconstruction schemes such as PPM and WENO5-Z\footnote{See Appendices~\ref{appendix:basic-framework} and \ref{appendix:reconstruction} for a brief introduction to the Godunov-type methods and high-order reconstruction schemes.}.

\subsection{Estimation of Numerical Dissipation}

In the next two sections, we present a general residual-based framework for the \textit{a posteriori} estimation of numerical dissipation with a specific focus on simulations of MHD turbulence. A key contribution of this framework is the ability to fully capture the anisotropy of numerical dissipation, thereby enabling a comprehensive characterization of numerical viscosity and resistivity in anisotropic MHD flows.

Based on the definitions in Appendix~\ref{appendix:basic-framework}, we rewrite the system of conservation laws compactly as
\begin{equation}
\mathcal{L}(u) = 0,
\end{equation}
where the operator $\mathcal{L}$ is defined as $\partial/\partial t + \nabla \cdot \mathbb{F}(\cdot)$, and $\mathbb{F}$ is the flux function. Let $\tilde{u}$ denote the exact solution, such that $\mathcal{L}(\tilde{u}) = 0$. With the computational domain partition $\Omega = \bigcup_{i}\Omega_i$ and discrete time steps $t^{(n)}\in \mathbb{R}_{\ge 0}$, let $h \coloneqq \{(\Omega_i,t^{(n)})\}$ be a discretization of the spatiotemporal domain $\Omega \times \mathbb{R}_{\ge 0}$. Within the framework of the finite volume method (FVM), we define a discretization operator $\mathcal{I}_h$ as
\begin{equation}
\mathcal{I}_h(u) \coloneqq \{ \overline{u}_{h,i}^{(n)}\} \equiv u_h.
\end{equation}
Furthermore, the numerical scheme can be defined as an operator $\mathcal{L}_h$ satisfying $\mathcal{L}_h(u_h) = 0$ with numerical solution $u_h$. An ``ideal'' numerical scheme without truncation error, denoted by $\mathcal{L}_h^{\mathrm{ideal}}$, satisfies by definition
\begin{equation}
\forall u,\  \mathcal{L}_h^{\mathrm{ideal}}(\mathcal{I}_h(u)) = \mathcal{I}_h(\mathcal{L}(u)).
\end{equation}
However, constructing such an ideal scheme in practice is generally infeasible. Instead, the actual numerical scheme yields a solution $u_h$ with
\begin{equation}
\mathcal{L}_h(u_h) = \mathcal{L}_h(\mathcal{I}_h(u'))  =  0,
\end{equation}
where $u'$ denotes a continuous reconstruction of the discrete numerical solution $u_h$.

In MEA, it is well-known that the continuous numerical solution $u'$ does not satisfy the original PDE $\mathcal{L}(u) = 0$, but rather a modified equation that includes the truncation errors \citep{Warming1974}:
\begin{equation}\label{eq:modified-pde}
\mathcal{L}(u') = \mathcal{D}^{\mathrm{num}}(u'),
\end{equation}
where the truncation term $\mathcal{D}^{\mathrm{num}}(u')$ is treated herein as numerical dissipation. Applying the definition of the ideal scheme $\mathcal{L}_h^{\mathrm{ideal}}$ to Equation~\eqref{eq:modified-pde} yields
\begin{equation}
\mathcal{L}_h^{\mathrm{ideal}}(u_h) = \mathcal{I}_h\left(\mathcal{D}^{\mathrm{num}}(u')\right)\equiv \mathcal{D}^{\mathrm{num}}_h(u_h).
\end{equation}
To estimate the numerical dissipation $\mathcal{D}^{\mathrm{num}}(u')$ in practice, we approximate the ideal scheme $\mathcal{L}_h^{\mathrm{ideal}}$ using a high-order scheme $\mathcal{L}_h^{\mathrm{HO}}$ (which possesses a higher order of accuracy than the original scheme $\mathcal{L}_h$). Consequently, the numerical dissipation can be evaluated \textit{a posteriori} solely from the numerical solution $u_h$, without relying on any additional \textit{a priori} assumptions, as
\begin{equation}
\mathcal{D}^{\mathrm{num}}_h (u_h)\approx\mathcal{L}_h^{\mathrm{HO}}(u_h).
\end{equation}

In this work, the modified equations of MHD momentum equation~\eqref{eq:momentum} and induction equation~\eqref{eq:induction} are employed to define the expressions of numerical viscosity and resistivity, respectively. Taking the magnetic induction equation as an illustrative example, the numerical resistive term $\bm{D}^{\mathrm{num}}_{\mathrm{res}}$ arises in the modified equation as:
\begin{equation}\label{eq:modified-induction-equation}
\frac{\partial \bm{B}}{\partial t} = \nabla \times (\bm{u}\times \bm{B}) + \eta \nabla^2\bm{B} + \bm{D}^{\mathrm{num}}_{\mathrm{res}}.
\end{equation}
In practice, the numerical resistive term can be evaluated \textit{a posteriori} by applying the high-order scheme $\mathcal{L}_h^{\mathrm{HO}}$ to the numerical solution $u_h$:
\begin{equation}\label{eq:numerical-resistive-term}
    \bm{D}^{\mathrm{num}}_{\mathrm{res}} = \left( \frac{\partial \bm{B}_h}{\partial t} \right)^{\mathrm{HO}}_h
    -
    \nabla^{\mathrm{HO}}_h \times (\bm{u}_h\times \bm{B}_h) - \eta \left(\nabla^2\right)^{\mathrm{HO}}_h\bm{B}_h,
\end{equation}
and the subscript $h$ is omitted hereafter for simplicity. Building upon the above definition of $\bm{D}^{\mathrm{num}}_{\mathrm{res}}$, the modified equation for the magnetic energy transport equation for the $i$-th component ($i=x,y,z$) is given by\footnote{A detailed derivation of the kinetic and magnetic energy transport equations is presented in Appendix~\ref{sec:transport}.}
\begin{equation}
    \frac{\partial}{\partial t}\left(\frac{B_i^2}{2}\right) = B_i [\nabla \times (\bm{u}\times \bm{B})]_i + \eta B_i \nabla^2 B_i + \mathscr{D}^{\mathrm{num}}_{\mathrm{res},i},
\end{equation}
where $\mathscr{D}^{\mathrm{num}}_{\mathrm{res},i} = B_i D^{\mathrm{num}}_{\mathrm{res},i}$ is the component-wise numerical resistive dissipation rate\footnote{To avoid ambiguity, the Einstein summation convention is not adopted throughout this paper, i.e., repeated indices do not imply summation.}, with $D^{\mathrm{num}}_{\mathrm{res},i}$ denoting the $i$-th component of $\bm{D}^{\mathrm{num}}_{\mathrm{res}}$. Following an analogous derivation for the momentum equation, the numerical viscous term $\bm{D}^{\mathrm{num}}_{\mathrm{vis}}$ can be obtained accordingly. Consequently, the total numerical viscous and resistive dissipation rates, denoted by
$\varepsilon^{\mathrm{num}}_{\mathrm{vis}}$ and $\varepsilon^{\mathrm{num}}_{\mathrm{res}}$,
respectively, can be evaluated by taking the inner products with the flow fields
(i.e., $\varepsilon = \sum_{i} \mathscr{D}_{i}$):
\begin{equation}
    \varepsilon^{\mathrm{num}}_{\mathrm{vis}} = \bm{u}\cdot \bm{D}^{\mathrm{num}}_{\mathrm{vis}}, \
    \varepsilon^{\mathrm{num}}_{\mathrm{res}} = \bm{B}\cdot \bm{D}^{\mathrm{num}}_{\mathrm{res}},
\end{equation}
consistent with the definition of numerical dissipation in \citet{Schranner2015} derived directly from the kinetic energy transport equation.

For simulations of anisotropic MHD turbulence, in particular, the component-wise estimation of the numerical dissipation terms, $D^{\mathrm{num}}_{\mathrm{vis},i}$ and $D^{\mathrm{num}}_{\mathrm{res},i}$, and the numerical dissipation rates, $\mathscr{D}^{\mathrm{num}}_{\mathrm{vis},i}$ and $\mathscr{D}^{\mathrm{num}}_{\mathrm{res},i}$, in the proposed framework offers a more comprehensive characterization of numerical dissipation than traditional methods, which are restricted to the total numerical dissipation rates $\varepsilon^{\mathrm{num}}_{\mathrm{vis}}$ and $\varepsilon^{\mathrm{num}}_{\mathrm{res}}$.

\subsection{High-order Schemes}\label{sec:high-order-schemes}

Accurate estimation of numerical dissipation relies on the construction of the high-order scheme $\mathcal{L}_h^{\mathrm{HO}}$ introduced in the previous section. Furthermore, fully developed turbulence is characterized by multi-scale structures spanning eddies from the energy-containing scales down to the dissipation scales. Consequently, the residual-based estimation of numerical dissipation requires a $\mathcal{L}_h^{\mathrm{HO}}$ with high spectral resolution. While spectral difference schemes provide perfect spectral resolution, they are prone to Gibbs oscillations near discontinuities and therefore are not suitable for compressible MHD turbulence. Accordingly, it is necessary to develop a high-order scheme with spectral-like resolution while retaining robust shock-capturing capability.

Here, we propose a novel targeted compact scheme with a multi-stencil discontinuity detector to implement the spatial derivative operators in $\mathcal{L}_h^{\mathrm{HO}}$, thereby achieving optimized spectral resolution with robust shock-capturing capability. For temporal derivatives, we employ Fornberg's finite difference formulas on arbitrarily spaced grids \citep{Fornberg1988} to accommodate the dynamic time steps determined by the CFL condition in MHD codes.

\subsubsection{Multi-stencil Discontinuity Detectors}\label{sec:msdd}

\begin{figure}[!tp]
    \centering
    \begin{subfigure}[b]{\linewidth}
        \centering
        \includegraphics[width=\linewidth]{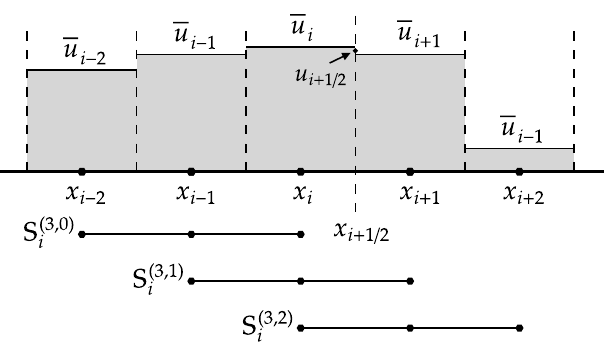}
        \subcaption{Substencil configuration within the five-point stencil $\mathrm{S}^{(5)}_i$ for WENO5-Z and TENO5-N schemes.}
        \label{fig:weno5}
    \end{subfigure}
    \begin{subfigure}[b]{\linewidth}
        \centering
        \includegraphics[width=\linewidth]{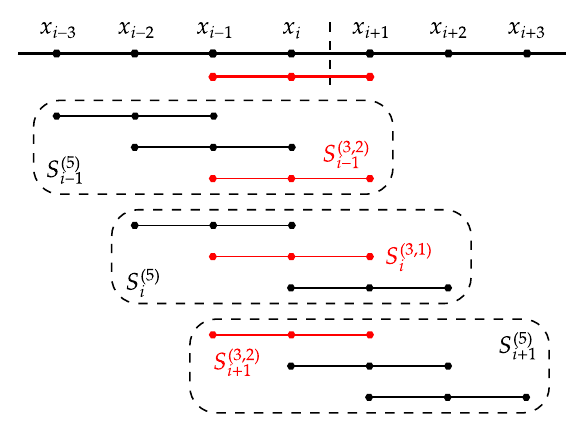}
        \subcaption{Stencil configuration for the multi-stencil discontinuity detector (MSDD). Each dashed box represents one five-point stencil with three associated three-point substencils.}
        \label{fig:msdd}
    \end{subfigure}
    \caption{Five-point stencils for~\subref{fig:weno5} the WENO5-Z and TENO5-N\footnote{Unlike the convention adopted in \citet{Fu2016, Fu2018}, where $\mathrm{S}^{(3,0)}_{i}$ denotes the substencil $\{x_{i-1}, x_i, x_{i+1}\}$, the present work directly adopts the WENO substencil configuration for the TENO schemes.} schemes and~\subref{fig:msdd} the multi-stencil discontinuity detector (MSDD). The three red substencils in Figure~\subref{fig:msdd}, $\mathrm{S}^{(3,2)}_{i-1}$, $\mathrm{S}^{(3,1)}_i$, and $\mathrm{S}^{(3,0)}_{i+1}$, represent the same three-point set $\{x_{i-1}, x_i, x_{i+1}\}$, and should in principle yield consistent $\delta$ values. In TENO-N, however, the discontinuity indicators associated with these substencils are evaluated independently, which could potentially yield inconsistent results. This motivates the unanimous criterion~\eqref{eq:criterion} adopted in the MSDD.}
\end{figure}

The high-order WENO \citep{Jiang1996, Shu2009} and TENO \citep{Fu2016, Fu2018, Fu2023} schemes are widely adopted in high-fidelity simulations of compressible turbulence due to their robust shock-capturing capabilities. In both schemes, the cell-interface value is reconstructed on a (e.g., five-point) stencil $\mathrm{S}^{(5)}_i$ using three substencils $\mathrm{S}^{(3,k)}_i$, as illustrated in Figure~\ref{fig:weno5}. The WENO schemes construct a weighted combination of candidate polynomials defined on the underlying substencils, using weights assigned according to their local smoothness. By comparison, the TENO-N scheme adopts a stencil-selection strategy using a discontinuity detector $\delta$ to categorize the substencils $\mathrm{S}^{(3,k)}_i$ into two categories: substencils crossed by discontinuities with $\delta=0$, and smooth substencils with $\delta=1$. A more systematic introduction to the WENO and TENO schemes is provided in Appendix~\ref{appendix:reconstruction}.

Unlike the shock-capturing schemes for numerical simulations, the high-order scheme $\mathcal{L}_h^{\mathrm{HO}}$ introduced for the \textit{a posteriori} estimation of numerical dissipation is not constrained by the stability requirements, which allows greater flexibility in designing a spectral-like scheme for $\mathcal{L}_h^{\mathrm{HO}}$. In addition, the TENO-N scheme may introduce a subtle inconsistency: Consider the three substencils $\mathrm{S}^{(3,2)}_{i-1}$, $\mathrm{S}^{(3,1)}_i$, and $\mathrm{S}^{(3,0)}_{i+1}$ from neighboring stencils $\mathrm{S}^{(5)}_{i-1}$, $\mathrm{S}^{(5)}_{i}$, and $\mathrm{S}^{(5)}_{i+1}$, respectively, as shown in Figure~\ref{fig:msdd}. These substencils all represent a common three-point set $\{x_{i-1}, x_{i}, x_{i+1}\}$, and thus should be consistently identified as either smooth or containing discontinuities. Yet, the three discontinuity indicators, $\delta^{(3,1)}_i$, $\delta^{(3,2)}_{i-1}$, and $\delta^{(3,0)}_{i+1}$, are determined using smoothness indicators from three distinct stencils, which could potentially yield inconsistent results. This inconsistency can result in misidentifying discontinuities in smooth regions and causing accuracy degeneration, especially in regions characterized by high local wavenumbers (e.g., small-scale turbulent eddies).

Motivated by this, we propose a more robust discontinuity detection strategy to resolve such inconsistencies. The core concept is to replace the criterion for detecting discontinuities in TENO-N with a unanimous voting strategy, where a substencil is identified as discontinuous if and only if all relevant $\delta$ values are $0$. Specifically, the condition for detecting a discontinuity within the central substencil $\mathrm{S}^{(3,1)}_i$ in TENO-N, i.e., $\delta^{(3,1)}_i = 0$, is replaced with a unanimous agreement among the three relevant $\delta$ values from neighboring stencils:
\begin{equation}\label{eq:criterion}
\delta^{(3,1)}_i= \delta^{(3,2)}_{i-1}= \delta^{(3,0)}_{i+1} = 0,
\end{equation}
as illustrated in Figure~\ref{fig:msdd}. Similarly, the criteria for detecting discontinuities within the substencil $\mathrm{S}^{(3,0)}_i$ and substencil $\mathrm{S}^{(3,2)}_i$ are replaced with $\delta ^{(3,0)}_i=\delta ^{(3,1)}_{i-1}= \delta ^{(3,2)}_{i-2} =0$ and $\delta^{(3, 2)}_i=\delta ^{(3,1)}_{i+1}= \delta ^{(3,0)}_{i+2} = 0$, respectively. Meanwhile, the criteria for identifying specific discontinuity configurations from the smoothness of individual substencils remain identical to those in the TENO-N scheme.

In the proposed strategy, confirming a specific discontinuity configuration requires satisfying nine conditions simultaneously, instead of only three as in the standard TENO-N scheme. This more stringent criterion effectively preserves the robust shock-capturing capability of TENO, while preventing potential accuracy degeneration induced by local high-wavenumber fluctuations. In this work, the novel discontinuity-detecting strategy is termed the multi-stencil discontinuity detector (MSDD), since it integrates multiple discontinuity indicators ($\delta_i, \delta_{i-1}, \delta_{i+1}$) across neighboring stencils to detect a consistent distribution of discontinuities. The resulting scheme, combining MSDD with the candidate reconstruction polynomials of TENO-N, is henceforth referred to as TENO-M (``M'' for ``multi-stencil'').

\begin{figure*}[!tp]
    \centering
    \begin{subfigure}[b]{\linewidth}
        \centering
        \includegraphics[width=0.96\linewidth]{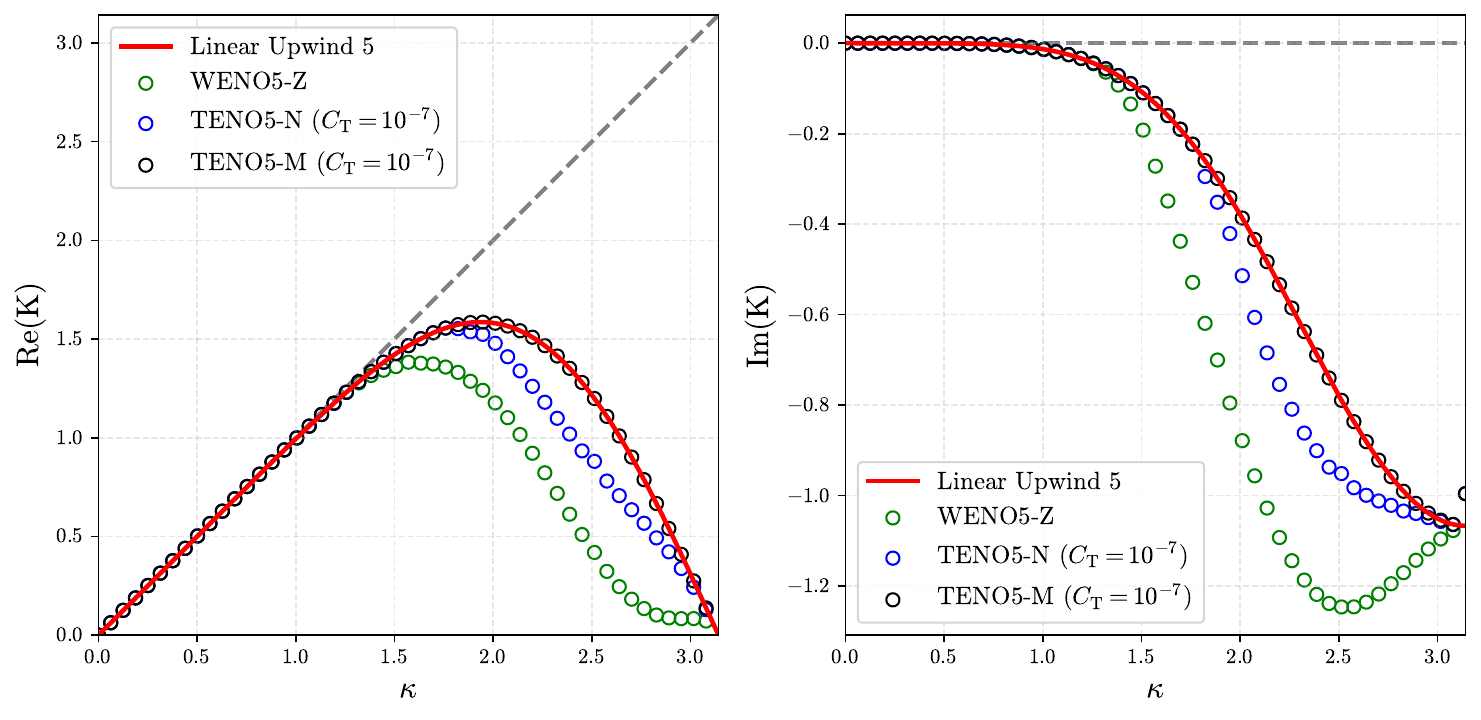}
        \subcaption{ADR analysis for various five-point reconstruction schemes. Compared with conventional WENO5-Z and TENO5-N schemes, the TENO5-M scheme proposed here effectively recovers the spectral properties of the underlying fifth-order linear upwind scheme (indicated by the solid red line).}
        \label{fig:adr-a}
    \end{subfigure}
    \begin{subfigure}[b]{\linewidth}
        \centering
        \includegraphics[width=0.96\linewidth]{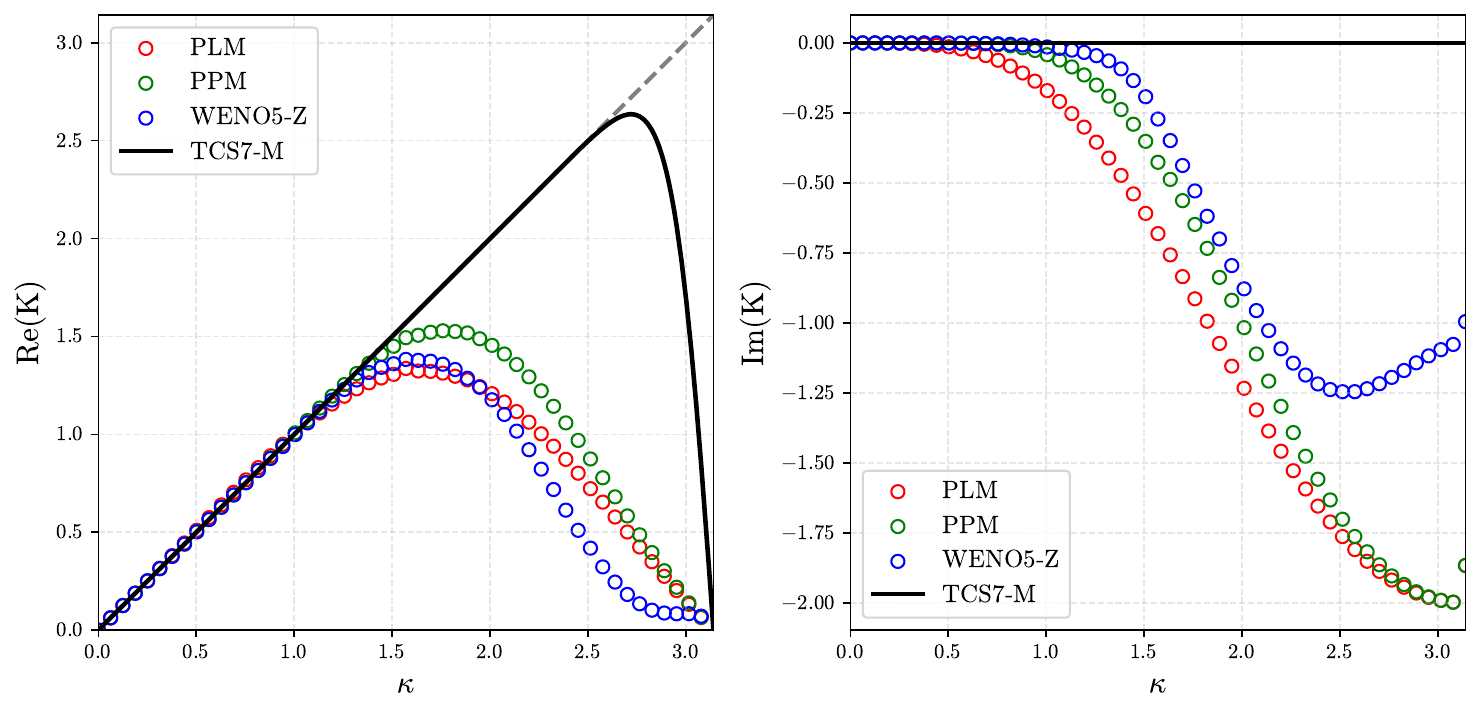}
        \subcaption{ADR analysis for the spectrally optimized TCS7-M scheme, compared with PLM, PPM, and WENO5-Z schemes commonly used in astrophysical MHD codes. The dispersion curve $\mathrm{Re}(\mathrm{K})-\kappa$ of TCS7-M scheme shows that its well-resolved wavenumbers reach up to $\kappa \approx 0.8\pi$, thereby exhibiting spectral-like resolution. Note that the TCS7-M scheme also introduces no dissipation error, i.e., $\mathrm{Im}(\mathrm{K})=0$, since it reduces to a symmetric central compact scheme in the smooth regions.}
        \label{fig:adr-b}
    \end{subfigure}
    \caption{Approximate dispersion relation (ADR) analysis for the proposed \subref{fig:adr-a} TENO5-M and \subref{fig:adr-b} TCS7-M schemes against selected conventional schemes: dispersion (left) and dissipation\footnote{Note that the dissipation shown here (i.e., the imaginary part of the modified wavenumber) is distinct from the numerical dissipation discussed in the main text.} (right) properties. $\kappa = 2\pi kL/N \in [0, \pi]$ represents the reduced wavenumber, with $\kappa=\pi$ corresponding to the Nyquist wavenumber of the numerical grid, and $\mathrm{K}$ denotes the modified wavenumber associated with the numerical schemes. Gray dashed lines indicating the exact (ideal) spectral properties without dispersion and dissipation errors are shown for reference.}
    \label{fig:adr}
\end{figure*}

As illustrated in Figure~\ref{fig:adr-a}, the approximate dispersion relation (ADR) analysis\footnote{An introduction to ADR analysis is provided in Appendix~\ref{sec:adr}.} \citep{Pirozzoli2006} of selected schemes shows that the proposed TENO-M scheme effectively recovers the spectral properties of the underlying linear scheme. In contrast, conventional WENO-Z and TENO-N schemes exhibit excessive numerical dispersion and dissipation, particularly in the high-wavenumber regimes.

\subsubsection{Spatial Derivatives in FVM} \label{sec:fvm-derivative}

For simplicity, consider an arbitrary function $u(x)$ discretized on a uniform one-dimensional Cartesian grid $\{x_i=i\Delta x\}$. Applying the fundamental theorem of calculus yields the relationship between the cell-averaged spatial derivatives and the reconstructed values at the cell interfaces $x_{i\pm 1/2}$:
\begin{equation}\label{eq:fvm-derivative}
\left.
\overline{\frac{\mathrm{d}u}{\mathrm{d}x}} \right|_i = \frac{1}{\Delta x} \left( u_{i+1/2} - u_{i-1/2}\right).
\end{equation}
Furthermore, \citet{Shu1989} established a fundamental duality between cell-centered derivatives in the finite difference method (FDM) and cell-averaged derivatives in FVM. Given an arbitrary function $u(x)$, define a function $h(x)$ implicitly via a sliding average:
\begin{equation}\label{eq:sliding-average}
\frac{1}{\Delta x}\int_{x-\Delta x/2}^{x+\Delta x/2} h(\xi)\ \mathrm{d}\xi = u(x).
\end{equation}
Then, substituting $x = x_i$ yields
\begin{equation}
\overline{h}_i = \frac{1}{\Delta x}\int_{x_{i-1/2}}^{x_{i+1/2}} h(x)\ \mathrm{d}x = u(x_i) = u_i.
\end{equation}
Differentiating Equation~\eqref{eq:sliding-average} with respect to $x$ and subsequently setting $x = x_i$ yields
\begin{equation}
\left. \frac{\mathrm{d}u(x)}{\mathrm{d}x} \right|_i = \frac{1}{\Delta x} \left[ h(x_{i+1/2}) - h(x_{i-1/2})\right].
\end{equation}
A direct comparison with Equation~\eqref{eq:fvm-derivative} reveals that calculating cell-centered derivatives from cell-centered variables (in FDM) is equivalent to calculating cell-averaged derivatives from cell-averaged variables (in FVM). This equivalence justifies the direct application of ADR analysis to FVM reconstruction schemes, as illustrated in Figure~\ref{fig:adr}.

While Equation~\eqref{eq:fvm-derivative}, combined with high-order reconstruction schemes, provides a method for constructing cell-averaged spatial derivatives in $\mathcal{L}_h^{\mathrm{HO}}$ for FVM, the treatment of nonlinear terms in the governing equations, e.g., the convective term $\nabla\cdot (\bm{u}\bm{u})$, requires special care: A fundamental restriction of FVM is that the cell average of a product is generally not equivalent to the product of cell averages \citep{Felker2018}, i.e., $\overline{u\cdot v} \neq \overline{u} \cdot \overline{v}$. In one-dimensional FVM, this can be resolved by multiplying the reconstructed cell-interface values directly. However, in three-dimensional FVM, the reconstructed interface values from cell-averaged variables represent face-averaged values $\overline{u}_{i+1/2,j,k}$ rather than face-centered values $u_{i+1/2,j,k}$, while $\overline{u_{i+1/2,j,k}v_{i+1/2,j,k}} \neq \overline{u}_{i+1/2,j,k} \overline{v}_{i+1/2,j,k}$. For accurate evaluation of the nonlinear terms, a high-order transformation from cell averages ($\overline{u}_{i,j,k}$) to cell centers ($u_{i,j,k}$) is required \citep{Felker2018}.

To resolve this, we perform a fourth-order truncated three-dimensional Taylor series expansion of the function $u(x,y,z)$ around the cell center $(x_i, y_j, z_k)$ and integrate over the cell volume $\Omega_{i,j,k}$. This establishes a direct relationship between the cell averages and the cell centers along with their derivatives (the subscripts ${i,j,k}$ and terms of order $\mathcal{O}(\Delta_i^4)$ and higher are neglected for brevity):
\begin{equation}\label{eq:ctr2avg4}
    \overline{u} = u
    + \frac{1}{24} \left[
    (\Delta x)^2 \frac{\partial^2 u}{\partial x^2}
    + (\Delta y)^2 \frac{\partial^2 u}{\partial y^2}
    + (\Delta z)^2 \frac{\partial^2 u}{\partial z^2}
    \right].
\end{equation}
To invert this relationship and express the cell centers $u_{i,j,k}$ in terms of the cell averages $\overline{u}_{i,j,k}$, we apply a second-order truncated expansion
\begin{equation}
\overline{(\cdot)}_{i,j,k} = (\cdot)_{i,j,k}
+ \mathcal{O}(\Delta_i^2)
\end{equation}
to the second-order partial derivatives (e.g., ${\partial^2 u}/{\partial x^2}$) in Equation~\eqref{eq:ctr2avg4}, yielding:
\begin{equation}
u = \overline{u}
- \frac{1}{24} \left[
  (\Delta x)^2\overline{\frac{\partial^2 u}{\partial x^2}}
+ (\Delta y)^2\overline{\frac{\partial^2 u}{\partial y^2}}
+ (\Delta z)^2\overline{\frac{\partial^2 u}{\partial z^2}}
\right],
\end{equation}
which serves as a transformation from cell-averaged variables to cell-centered variables with fourth-order accuracy. Such a transformation is essential for preserving the formal accuracy of high-order Godunov-type schemes, which would otherwise be restricted to second-order spatial accuracy. Furthermore, a more sophisticated sixth-order transformation is derived in Appendix~\ref{sec:avg2ctr6} and employed in the construction of the high-order operator $\mathcal{L}_h^{\mathrm{HO}}$ in the proposed framework for the \textit{a posteriori} estimation of numerical dissipation for Godunov MHD codes.

To the best of our knowledge, among various astrophysical Godunov codes, only \texttt{Athena++} incorporates such a high-order average-to-center transformation \citep{Felker2018}\footnote{This algorithm is implemented in \texttt{src/eos/eos\_high\_order.cpp} of the \texttt{Athena++} repository (\url{https://github.com/PrincetonUniversity/athena}, commit \texttt{ed4d1e3}).}. However, its application is restricted to pure hydrodynamic simulations without magnetic fields and to specific reconstruction schemes.

\subsubsection{Targeted Compact Schemes}\label{sec:tcs}

Traditional (explicit) finite difference schemes often require large stencils to achieve high-order accuracy and high spectral resolution, where the derivative $u_i'$ is explicitly expressed as a function of cell-centered variables from a $(q+r+1)$-point stencil $\{u_{i-q}, \dots, u_{i+r}\}$. Compact finite difference schemes, in contrast, generalize this procedure by including the derivatives at neighboring grid points, leading to an implicit formulation. For example, a central compact scheme can be generally written as \citep{Lele1992}:
\begin{equation}\label{eq:central-compact-scheme}
\begin{aligned}
& \beta u'_{i-2} + \alpha u'_{i-1} + u'_i + \alpha u'_{i+1} + \beta u'_{i+2} = \\
& c \frac{u_{i+3} - u_{i-3}}{6\Delta x} + b \frac{u_{i+2} - u_{i-2}}{4\Delta x} + a \frac{u_{i+1} - u_{i-1}}{2\Delta x},
\end{aligned}
\end{equation}
where $u_i$ and $u_i'$ denote the cell-centered variable and its derivative at $x_i$, respectively. Applying Equation~\eqref{eq:central-compact-scheme} to a grid of $N$ points (with periodic boundary conditions, for instance) results in a linear system of $N$ equations for all the unknown derivatives $u_i'$. Solving this system (typically a quasi-tridiagonal or quasi-pentadiagonal linear system) yields the derivative values at all grid points. The above central compact scheme can achieve up to tenth-order accuracy while maintaining spectral-like resolution.

However, implicit central compact schemes are also prone to spurious oscillations in the vicinity of discontinuities, similar to explicit central difference schemes. To resolve this, \citet{Jiang2001} proposed the weighted compact scheme (WCS), in which Equation~\eqref{eq:central-compact-scheme} is replaced by a linear combination of candidate equations with the nonlinear weighting strategy of the WENO schemes (see Equation~\eqref{eq:nonlinear-weights}). Each candidate equation is either centered or biased so as to avoid a specific discontinuity distribution. Consequently, the scheme retains the high spectral resolution of central compact schemes in smooth regions while maintaining robust shock-capturing capability near discontinuities.

Nevertheless, WCS still introduces excessive numerical dissipation in the high-wavenumber regime, as does WENO. In this work, the WENO weighting strategy in WCS is replaced by the MSDD proposed in Section~\ref{sec:msdd}, leading to a novel targeted compact scheme with MSDD (TCS-M). In TCS-M, Equation~\eqref{eq:central-compact-scheme} is generalized to
\begin{equation}
    \sum_{l=-2}^2 \alpha_l u'_{i+l} = \frac{1}{\Delta x}\sum_{l=-3}^3 a_{l} u_{i+l},
\end{equation}
where the coefficients are determined according to the distribution of discontinuities within the stencil, as identified by the MSDD. The ADR analysis in Figure~\ref{fig:adr-b} shows that the proposed spectral-like TCS7-M scheme can effectively resolve wavenumbers up to $\kappa \approx 0.8\pi$. The complete set of candidate coefficients for all discontinuity configurations in TCS7-M is provided in Appendix~\ref{appendix:tcs}, which also describes the algorithm used to solve the sparse linear system associated with the compact schemes.

\subsubsection{High-order Time Derivatives}

In MHD simulations, the time step $\Delta t$ is typically determined dynamically by the CFL condition~\eqref{eq:cfl}. Consequently, conventional high-order finite difference schemes designed for uniform grids are ill-suited for implementing the $(\partial/\partial t)^{\mathrm{HO}}$ operator within the high-order scheme $\mathcal{L}_h^{\mathrm{HO}}$. To address this, we employ Fornberg's finite difference formulas on arbitrarily spaced grids \citep{Fornberg1988} as the high-order algorithm for temporal derivatives.

To estimate the time derivative at a given $t^{(n)}$, the simulation must be configured to output a sequence of consecutive data files at $t^{(n+k)}$ during runtime, where $k=-K, \cdots, K$ and $K$ is a positive integer. Using Fornberg's formulas, the derivative $(\partial u/\partial t)^{(n)}$ is approximated as:
\begin{equation}
\left(\frac{\partial u}{\partial t}\right)^{(n)} \approx \sum_{\nu=0}^{2K} \delta^1_{2K,\nu}u^{(n-K+\nu)},
\end{equation}
where $\delta^m_{n,\nu}$ denotes the finite difference weights determined by the time sequence $\{t^{(n+k)}\}$. In general, this approximation achieves an order of accuracy of $2K$. The specific recursive algorithm for the computation of the coefficients $\delta^m_{n,\nu}$ is detailed in \citet{Fornberg1988}.

\section{Properties of Numerical Dissipation} \label{sec:properties}

In this section, the proposed framework is applied to simulations of Alfv\'enic MHD turbulence, turbulent small-scale dynamos, and turbulent MRI-driven dynamos to systematically characterize the statistical properties of numerical dissipation in MHD turbulence simulations.

\subsection{Anisotropy of Numerical Dissipation}\label{sec:anisotropic}

As canonical examples of anisotropic turbulence, Alfv\'enic MHD turbulence and turbulent MRI-driven dynamos are considered here to demonstrate the capability of the proposed framework to characterize the anisotropy of numerical dissipation. We begin with Alfv\'enic MHD turbulence and perform driven turbulence simulations using \texttt{AthenaPK}, a Godunov MHD code that incorporates the generalized Lagrangian multiplier method for divergence cleaning of the magnetic field. Unless specified otherwise, the simulations adopt the PLM spatial reconstruction, van Leer time integrator with a CFL number of $0.3$, HLLD Riemann solver, and an adiabatic EoS with an adiabatic index of $\gamma=5/3$.

The Alfv\'enic MHD turbulence is driven by a purely solenoidal and isotropic driving force with a forcing spectrum proportional to $(k/k_{\mathrm{peak}})^2\left[2 - (k/k_{\mathrm{peak}})^2\right]$. The peak forcing wavenumber is set to $k_{\mathrm{peak}}=2.0$ (normalized to the box size $L=1.0$), corresponding to a peak energy injection scale of $L_{\mathrm{inj}}=0.5$. The stochastic driving force evolves in time following an OU process with an autocorrelation time of $T_{\mathrm{corr}}=1.0$, and is normalized at each time step to maintain a constant root-mean-square acceleration of $a_{\mathrm{rms}}=1.0$. The simulation is initialized with uniform initial conditions of\footnote{A relatively large initial pressure $p_0$ is adopted here, as our preliminary tests indicated that a smaller value could lead to the generation of unphysical negative pressures.} $\rho_0 = 1.0$, $\bm{u}_0 = 0$, $p_0 = 10.0$, and $\bm{B}_0 = 0.5\bm{e}_x$, and evolved for at least five large-eddy turnover times to reach a statistically stationary turbulent state. Subsequently, we configured the simulation to output at least nine consecutive data files containing both primitive variables and the driving force for the analysis of numerical dissipation. The numerical dissipation terms (e.g., the numerical resistive term $\bm{D}^{\mathrm{num}}_{\mathrm{res}}$ in Equation~\eqref{eq:numerical-resistive-term}) were then estimated using the proposed framework detailed in Section~\ref{sec:methods}. Moreover, two different schemes were employed for the specific implementation of spatial derivatives in $\mathcal{L}_h^{\mathrm{HO}}$ introduced in Section~\ref{sec:high-order-schemes}, namely TENO-M in Section~\ref{sec:msdd} and TCS-M in Section~\ref{sec:tcs}. The results obtained with these two schemes are highly consistent, indicating that both schemes are sufficiently accurate for the \textit{a posteriori} characterization of numerical dissipation in the MHD turbulence simulations considered here. For simulations with higher-order reconstructions, e.g., the WENO5-Z scheme, TENO-M may not provide sufficient spectral resolution for the residual estimation, and the spectral-like TCS-M scheme is required instead.

\begin{figure*}[t!]
    \centering
    \begin{subfigure}[b]{\linewidth}
    \includegraphics[width=\linewidth]{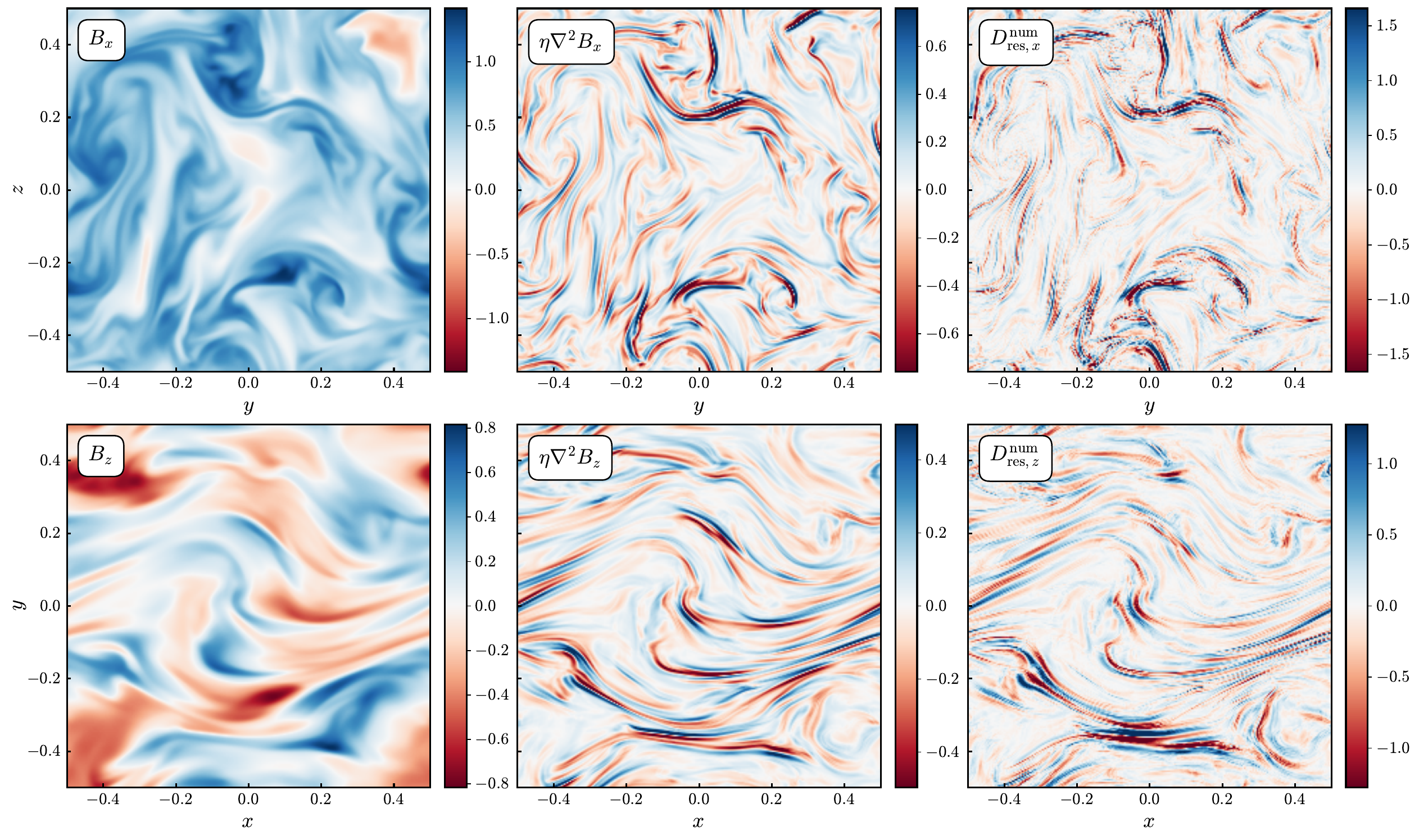}
        \caption{Components of magnetic field $\bm{B}$, physical resistive term $\eta\nabla^2\bm{B}$, and numerical resistive term $\bm{D}^{\mathrm{num}}_{\mathrm{res}}$.}
        \label{fig:Bx-slice-res}
    \end{subfigure}
    \begin{subfigure}[b]{\linewidth}
    \includegraphics[width=\linewidth]{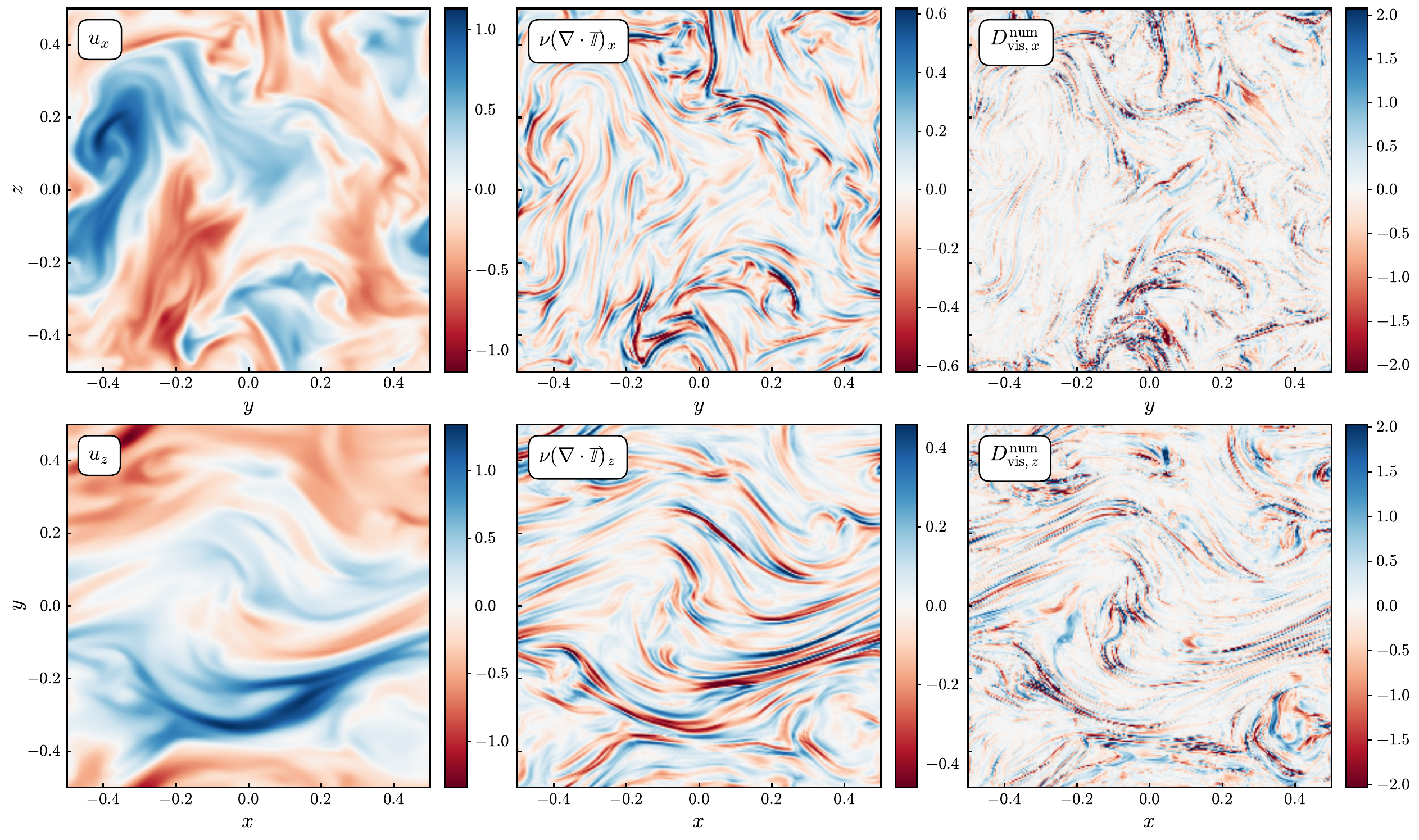}
        \caption{Components of velocity $\bm{u}$, physical viscous term $\nu \nabla \cdot \mathbb{T}$, and numerical viscous term $\bm{D}^{\mathrm{num}}_{\mathrm{vis}}$.}
        \label{fig:Bx-slice-vis}
    \end{subfigure}
    \caption{Two-dimensional slices of physical and numerical dissipation terms in a simulation of Alfv\'enic MHD turbulence. The top rows show the $y$--$z$ slices of $x$-components at the $x=0$ plane, and the bottom rows show the $x$--$y$ slices of $z$-components at the $z=0$ plane. The background magnetic field leads to the formation of anisotropic structures elongated along the $x$-direction.}
    \label{fig:Bx-slice}
\end{figure*}

Figure~\ref{fig:Bx-slice} illustrates the spatial distribution of both physical and numerical dissipation terms from a $256^3$ simulation of weakly compressible, strong Alfv\'enic MHD turbulence with sonic Mach number $\mathcal{M}_{\mathrm{s}}=0.12$, Alfv\'en Mach number $\mathcal{M}_{\mathrm{A}}=0.9$, and physical dissipation coefficients $\nu=\eta=10^{-4}$. In this simulation, the physical viscous and resistive terms $\bm{D}^{\mathrm{phy}}_{\mathrm{vis}}$ and $\bm{D}^{\mathrm{phy}}_{\mathrm{res}}$ are concentrated in filamentary, spatially coherent structures associated with large gradients in the velocity and magnetic fields, and they exhibit noticeable anisotropy along the $x$-direction due to the presence of the background magnetic field. By contrast, the numerical dissipation terms $\bm{D}^{\mathrm{num}}_{\mathrm{vis}}$ and $\bm{D}^{\mathrm{num}}_{\mathrm{res}}$ remain spatially correlated with their physical counterparts, indicating that numerical dissipation is not distributed randomly but rather follows the large-scale structure of the physical dissipation. In particular, the numerical resistive terms reproduce the overall morphology of the physical resistive terms, whereas the numerical viscous terms are more intermittent and exhibit more evident small-scale nonphysical oscillations.

\begin{figure*}[t!]
    \centering
    \includegraphics[width=\linewidth]{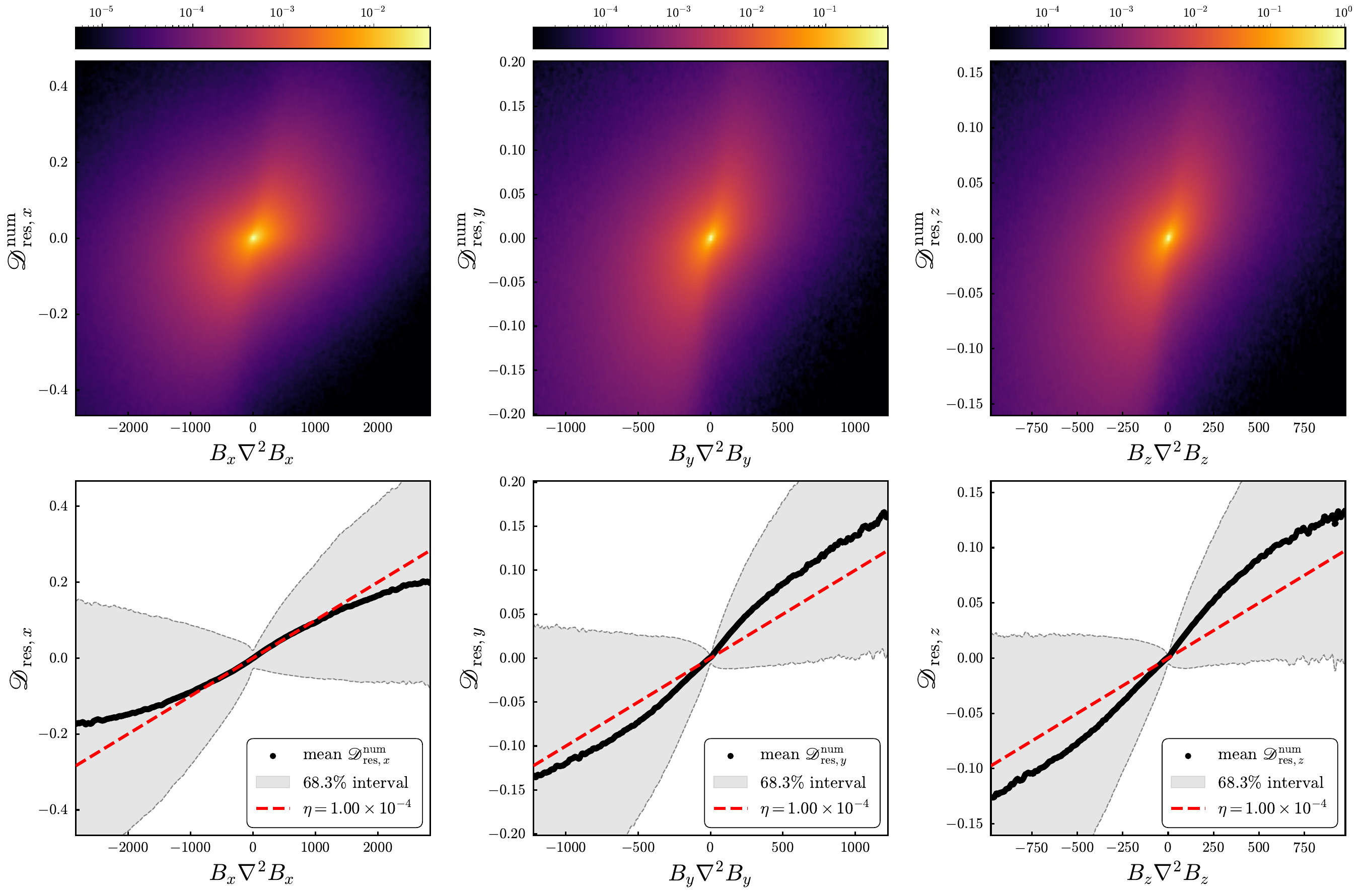}
    \caption{Top row: joint probability density functions (JPDFs) of the component-wise numerical resistive dissipation rate $\mathscr{D}^{\mathrm{num}}_{\mathrm{res},i}$ and $B_i\nabla^2 B_i$ in a simulation of Alfv\'enic MHD turbulence with the background magnetic field in the $x$-direction. Note that $B_i\nabla^2 B_i$ is the component-wise physical resistive dissipation rate $\mathscr{D}^{\mathrm{phy}}_{\mathrm{res},i}=\eta B_i\nabla^2 B_i$ without the factor $\eta$. Bottom row: conditional means of $\mathscr{D}^{\mathrm{num}}_{\mathrm{res},i}$ at fixed $B_i\nabla^2 B_i$ (black dots), along with the $68.3\%$ interval (gray shaded areas) and red dashed lines indicating the physical resistivity $\eta$ for reference. The conditional means of $\mathscr{D}^{\mathrm{num}}_{\mathrm{res},i}$ demonstrate that the numerical resistive dissipation is intrinsically anisotropic: the $x$-component is lower than the $y$- and $z$-components, reflecting the anisotropy of the turbulence.}
    \label{fig:Bx-hist-res}
\end{figure*}

Figure~\ref{fig:Bx-hist-res} shows the joint probability density functions (JPDFs) of the component-wise numerical resistive dissipation rate $\mathscr{D}^{\mathrm{num}}_{\mathrm{res},i}$ and $B_i\nabla^2 B_i$ for the $i=x,y,z$ components from the simulation described above, where $B_i\nabla^2 B_i$ is the component-wise physical resistive dissipation rate $\mathscr{D}^{\mathrm{phy}}_{\mathrm{res},i}=\eta B_i\nabla^2 B_i$ without the factor $\eta$. The continuous JPDFs are estimated from the discrete data using kernel density estimation (KDE) with the Python package \texttt{KDEpy} \citep{kdepy}. KDE is a non-parametric method for estimating a probability density function (PDF) from discrete data points, where a Gaussian kernel is placed on each data point with bandwidth estimated using Silverman's rule \citep{Silverman2018}, and the resulting continuous PDF is given by the sum of all kernel functions. The JPDFs reveal a positive, albeit weak, correlation between the component-wise numerical resistive dissipation rates $\mathscr{D}^{\mathrm{num}}_{\mathrm{res},i}$ and their physical counterparts $\mathscr{D}^{\mathrm{phy}}_{\mathrm{res},i}$. This weak correlation indicates that numerical dissipation cannot be simply interpreted as additional physical-like dissipation terms. Moreover, the conditional means of the component-wise dissipation rates, i.e., the mean values of $\mathscr{D}^{\mathrm{num}}_{\mathrm{res},i}$ at fixed values of $B_i\nabla^2 B_i$, show that the numerical dissipation is intrinsically anisotropic, with a clear discrepancy between the $x$-component and the $y$- and $z$-components, reflecting the anisotropy of the turbulence.

MRI-driven turbulence represents another prominent example of highly anisotropic turbulence in astrophysics. In differentially rotating disks, the initial magnetic field is exponentially amplified by the magnetorotational instability (MRI), with subsequent parasitic instabilities, including Kelvin-Helmholtz and tearing-mode instabilities \citep{Goodman1994}, disrupting the MRI channel flows, ultimately leading to a fully developed turbulent state. In local shearing-box simulations, the combination of the background shear in the $x$--$y$ plane and the magnetic field induces strong anisotropy in the MRI-driven turbulence. Therefore, it provides a representative example for demonstrating the capability of the proposed framework to characterize highly anisotropic numerical dissipation.

In this study, we performed shearing-box simulations of turbulent MRI-driven dynamos using \texttt{AthenaK}, a Godunov MHD code with constrained transport to enforce the divergence-free condition of the magnetic field. It is worth noting that in the orbital advection implementation within \texttt{AthenaK}, the physical viscous term is expressed in terms of the velocity fluctuation $\bm{v}$ instead of the total velocity $\bm{u}$ (i.e., with the background shear flow subtracted, see Section~\ref{sec:mri-driven-turbulence} for details), thereby effectively neglecting the contribution of the background shear to the viscous stress. Consistent with this approximation, a similar treatment is adopted in Equation~\eqref{eq:mri-momentum} for the implementation in the proposed framework. For the numerical setup in this section, we employed the PLM spatial reconstruction, Heun's time integrator with a CFL number of $0.4$, and the HLLD Riemann solver. Additionally, we adopted an isothermal EoS with a constant sound speed of $c_{\mathrm{s}}=1.0$ and a local angular velocity of $\Omega_{0}=1.0$, yielding a characteristic scale height of $H=c_{\mathrm{s}}/\Omega_{0}=1.0$. The system was evolved for at least thirty orbital periods ($T_{\mathrm{orb}}=2\pi/\Omega_{0}$) to reach a statistically stationary turbulent state. Subsequently, we configured the simulation to output a sequence of consecutive data files containing all primitive variables for the \textit{a posteriori} analysis of numerical dissipation. In addition, as introduced in Section~\ref{sec:mri-driven-turbulence}, the background shear flow $\bm{u}_0=-q\Omega_{0} x\,\bm{e}_y$ implies a relative shearing velocity between the two radial ($x$-direction) boundaries of $\Delta u_0=q\Omega_{0} L_x$. Consequently, the shear-periodic boundary condition in the $x$-direction becomes strictly periodic only at time intervals of $\Delta t_0=L_y/\Delta u_0=0.212~T_{\mathrm{orb}}$. Therefore, the output data files were configured only at time steps closest to integer multiples of $\Delta t_0$, yielding snapshots with approximate periodic boundary conditions that can thus be consistently used for spectral analysis and the estimation of numerical dissipation.

\begin{figure*}[t!]
    \centering
    \begin{subfigure}[b]{\linewidth}
    \includegraphics[width=\linewidth]{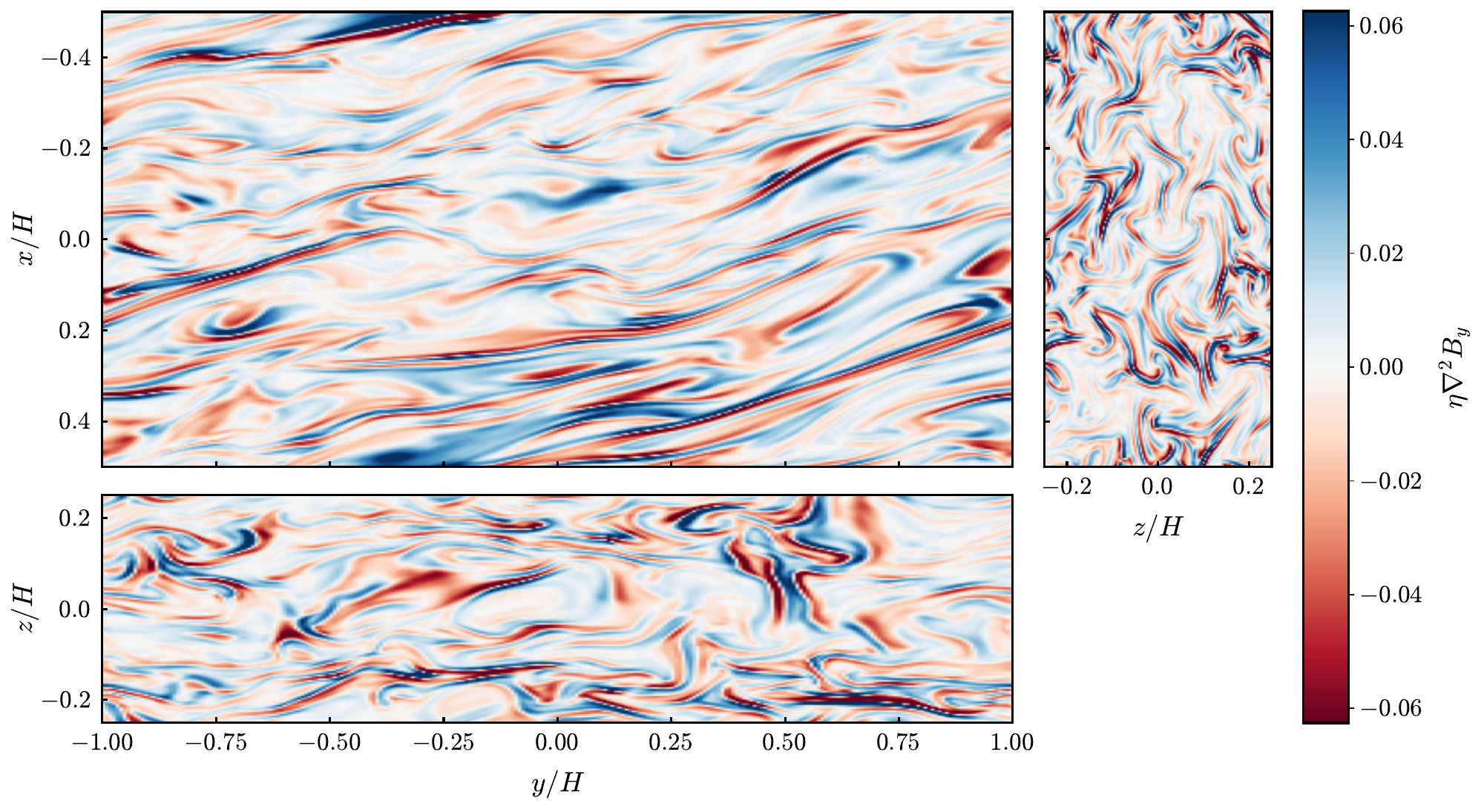}
        \caption{$y$-component of the physical resistive term $\eta\nabla^2 \bm{B}$.}
        \label{fig:mri-phy-res-y}
    \end{subfigure}
    \begin{subfigure}[b]{\linewidth}
        \includegraphics[width=\linewidth]{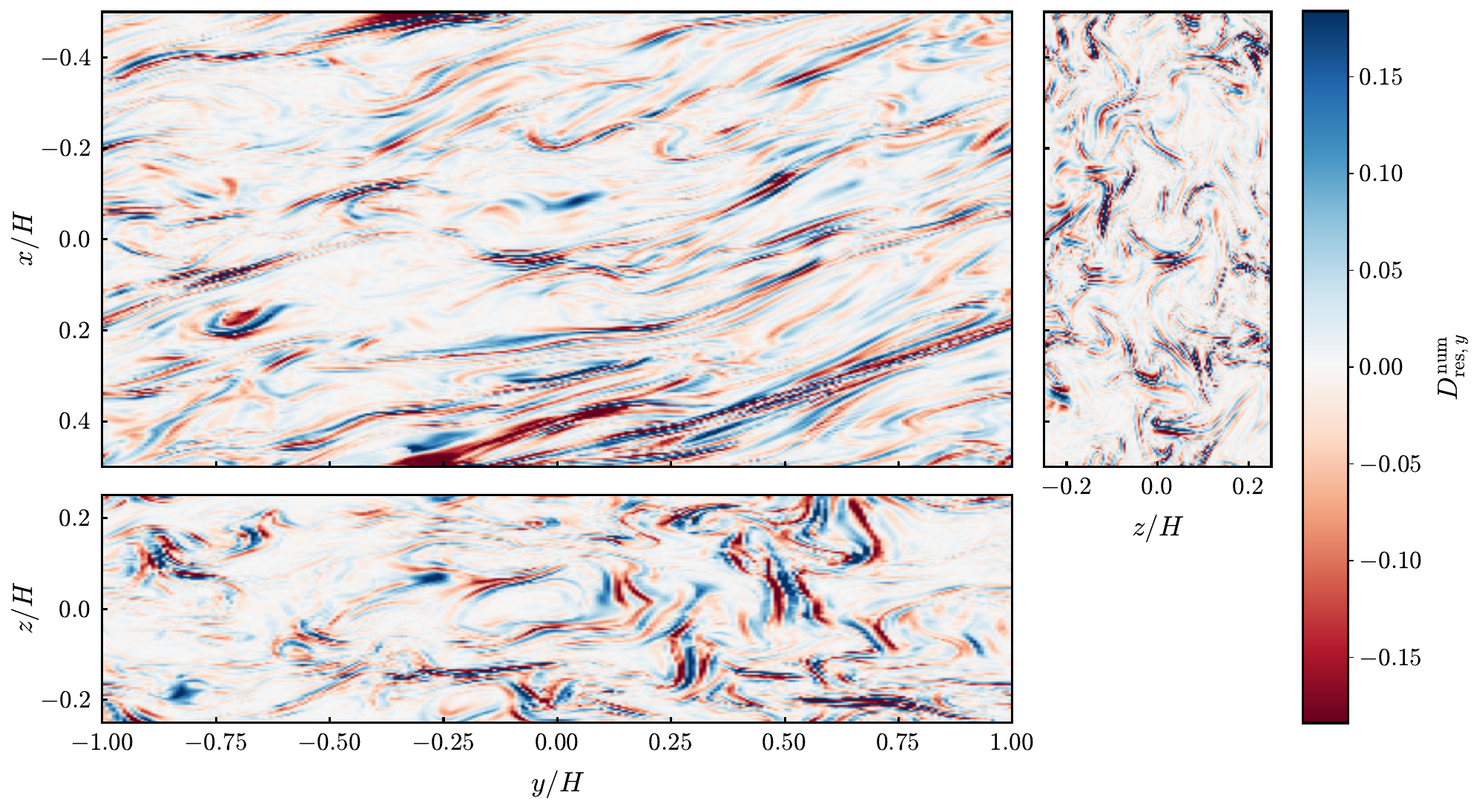}
        \caption{$y$-component of the numerical resistive term $\bm{D}^{\mathrm{num}}_{\mathrm{res}}$.}
        \label{fig:mri-num-res-y}
    \end{subfigure}
    \caption{Two-dimensional slices of physical and numerical resistive terms in a shearing-box simulation of a turbulent MRI-driven dynamo with $\mathrm{Re}=10^4$ and $\mathrm{Pm}=10$. The top-left panels show the $x$--$y$ slices in the $z=0$ plane, the bottom-left panels the $y$--$z$ slices in the $x=0$ plane, and the top-right panels the $x$--$z$ slices in the $y=0$ plane. The filamentary structures elongated primarily along the $y$ direction indicate strong anisotropy of the turbulence.}
    \label{fig:mri-slice}
\end{figure*}

Figure~\ref{fig:mri-slice} compares the physical and numerical resistive terms in a turbulent MRI-driven dynamo from a shearing-box simulation with resolution $256\times256\times128$, Reynolds number $\mathrm{Re}=10^4$, and magnetic Prandtl number $\mathrm{Pm}=10$. In the $x$--$y$ and $y$--$z$ slices, the $y$-component of the physical resistive term $D_{\mathrm{res},y}^{\mathrm{phy}}=\eta\nabla^2 B_y$ is organized into thin filamentary structures that are stretched predominantly along the $y$-direction by the background shear, reflecting the strong anisotropy of the turbulence. The $y$-component of the numerical resistive term $D_{\mathrm{res},y}^{\mathrm{num}}$ follows the same anisotropic structure, showing that numerical dissipation is primarily concentrated in the same regions where physical resistive dissipation is active. Meanwhile, the numerical resistive term is more spatially intermittent, containing thinner and more localized filaments, especially near the grid scale. Overall, this behavior is qualitatively consistent with that of Alfv\'enic turbulence shown in Figure~\ref{fig:Bx-slice}: the numerical dissipation reproduces the large-scale structure of the physical dissipation, while differences become more evident at small scales.

\begin{figure*}[t!]
    \centering
    \includegraphics[width=\linewidth]{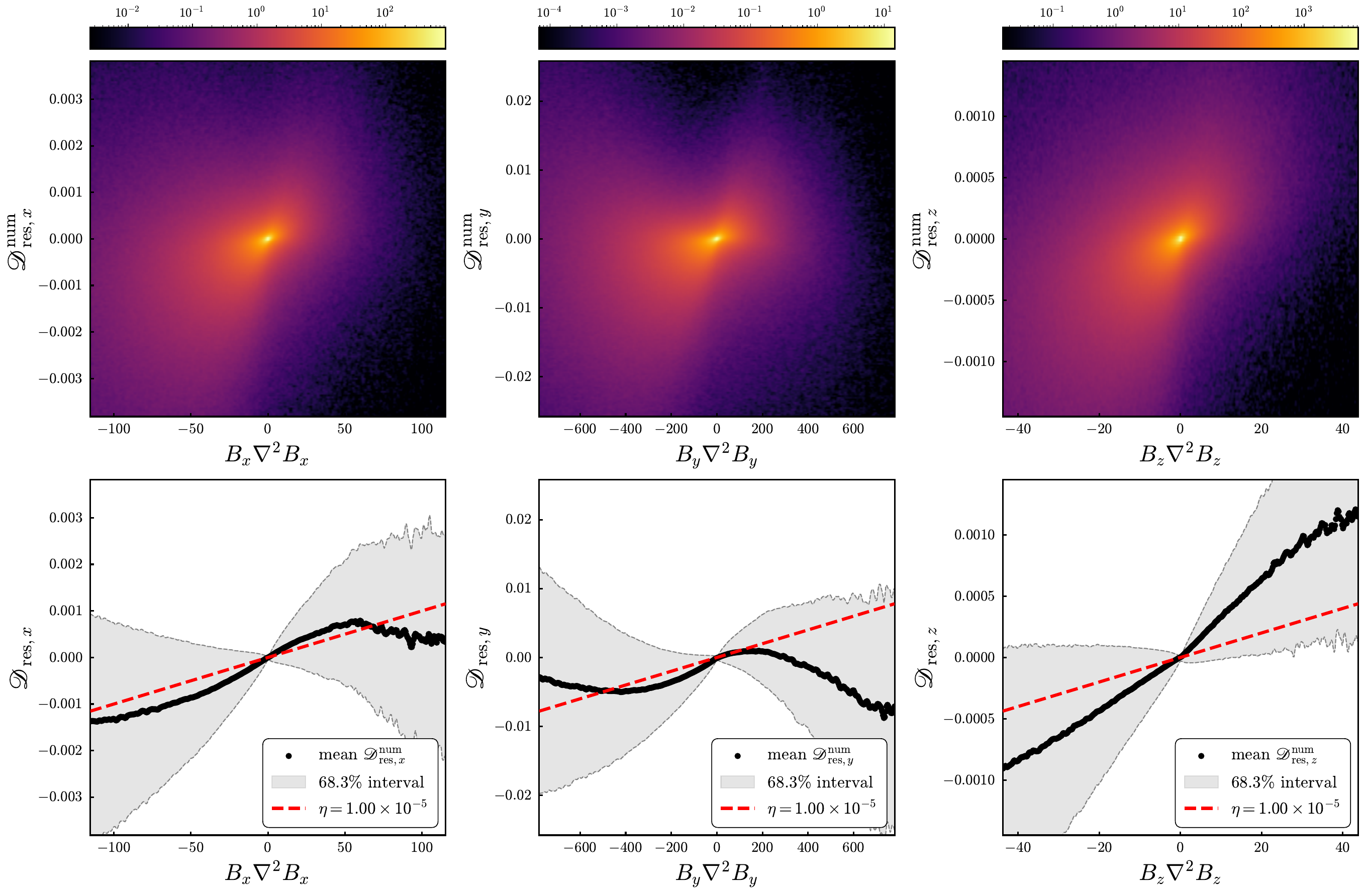}
    \caption{Same as Figure~\ref{fig:Bx-hist-res}, but for a shearing-box simulation of a turbulent MRI-driven dynamo with $\mathrm{Re}=10^4$ and $\mathrm{Pm}=10$. The conditional means of $\mathscr{D}^{\mathrm{num}}_{\mathrm{res},i}$ clearly demonstrate that the numerical resistive dissipation is highly anisotropic and significantly deviates from the behavior expected for physical-like resistive dissipation.}
    \label{fig:mri-hist}
\end{figure*}

Similar to Figure~\ref{fig:Bx-hist-res}, Figure~\ref{fig:mri-hist} presents the JPDFs of the component-wise numerical resistive dissipation rate $\mathscr{D}^{\mathrm{num}}_{\mathrm{res},i}$ and $B_i\nabla^2 B_i$ for the shearing-box simulation of the MRI-driven dynamo discussed above. The conditional means of the component-wise dissipation rates $\mathscr{D}^{\mathrm{num}}_{\mathrm{res},i}$ indicate strong anisotropy in the numerical resistive dissipation. Moreover, the $x$- and $y$-components exhibit the most pronounced departures from a physical-like resistivity.

\subsection{Spectral Analysis}\label{sec:spectral-analysis}

The apparent nonphysical behavior of the numerical dissipation suggests that its effects cannot be fully characterized by simple effective coefficients. It is therefore necessary to understand how the numerical energy dissipation is distributed across spatial scales and, in particular, how it shapes the turbulent cascade. To address this, we next examine the spectral properties of the numerical dissipation and its impact on the kinetic and magnetic energy spectra.

The definition of turbulent energy spectra can be derived from the Parseval-Plancherel identity \citep{Plancherel1910}: for any real-valued vector fields $\bm{g}(\bm{r})$ and $\bm{h}(\bm{r})$, their Fourier transforms (denoted by $\mathscr{F}$) satisfy
\begin{equation}\label{eq:parseval-plancherel}
    \int \bm{g}(\bm{r})\cdot \bm{h}(\bm{r})\ \mathrm{d}^3 \bm{r} =
    \int \tilde{\bm{g}}^*(\bm{k})\cdot \tilde{\bm{h}}(\bm{k})\ \mathrm{d}^3 \bm{k},
\end{equation}
where $\tilde{g}_i(\bm{k}) = \mathscr{F}[g_i(\bm{r})],\ \tilde{h}_i(\bm{k}) =\mathscr{F}[h_i(\bm{r})]$, and the superscript $^*$ denotes the complex conjugate. Specifically, setting $\bm{g} = \bm{h} = \bm{B}/\sqrt{2}$ in Equation~\eqref{eq:parseval-plancherel} yields the definition of magnetic energy spectrum
\begin{equation}
    E_{\mathrm{mag}}(\bm{k}) = \frac{1}{2}\left|\tilde{\bm{B}}(\bm{k})\right|^2, \tilde{B}_i(\bm{k}) = \mathscr{F}[B_i(\bm{r})].
\end{equation}
Similarly, introducing the density-weighted velocity
$\bm{w} = \sqrt{\rho}\bm{u}$ and setting
$\bm{g}=\bm{h}=\bm{w}/\sqrt{2}$ in Equation~\eqref{eq:parseval-plancherel} yields the definition of kinetic energy spectrum \citep{Kida1990}
\begin{equation}
    E_{\mathrm{kin}}(\bm{k}) = \frac{1}{2}\left|\tilde{\bm{w}}(\bm{k})\right|^2, \tilde{w}_i(\bm{k}) = \mathscr{F}[w_i(\bm{r})].
\end{equation}
Furthermore, the viscous and resistive dissipation rates generally take the form of dot products between two vector fields (see Appendix~\ref{sec:transport} for details), as
\begin{equation}
    \varepsilon_{\mathrm{vis}} = \bm{u}\cdot \bm{D}_{\mathrm{vis}}, \
    \varepsilon_{\mathrm{res}} = \bm{B}\cdot \bm{D}_{\mathrm{res}},
\end{equation}
which directly leads to the definitions of the viscous and resistive dissipation spectra as
\begin{equation}
    \begin{aligned}
    \varepsilon_{\mathrm{vis}}(\bm{k})& = \Re\left[\tilde{\bm{u}}^*(\bm{k})\cdot \tilde{\bm{D}}_{\mathrm{vis}}(\bm{k})\right], \\
    \varepsilon_{\mathrm{res}}(\bm{k})& = \Re\left[\tilde{\bm{B}}^*(\bm{k})\cdot \tilde{\bm{D}}_{\mathrm{res}}(\bm{k})\right],
    \end{aligned}
\end{equation}
where $\Re$ denotes the real part. Specifically, the numerical and physical dissipation spectra are defined with corresponding dissipation terms, and are denoted by $\varepsilon^{\mathrm{num}}_{\mathrm{vis}}(\bm{k})$, $\varepsilon^{\mathrm{phy}}_{\mathrm{vis}}(\bm{k})$, $\varepsilon^{\mathrm{num}}_{\mathrm{res}}(\bm{k})$, and $\varepsilon^{\mathrm{phy}}_{\mathrm{res}}(\bm{k})$, respectively.

In this section, we focus on the analysis of the spectral properties of numerical dissipation for the saturated phase in simulations of turbulent SSDs. The initial ZNF magnetic field configuration adopted in typical SSD simulations ensures that the saturated phase of the dynamo is statistically isotropic, thereby providing an example of isotropic MHD turbulence. We perform driven turbulence simulations with this ZNF configuration using the \texttt{AthenaPK} code. To compare the characteristics of numerical dissipation across different numerical schemes, both the PLM reconstruction (coupled with van Leer time integrator) and the PPM reconstruction (with SSP-RK3 time integrator) are employed. The driving force adopted here follows the same setup as that for the Alfv\'enic MHD turbulence, while the initial conditions are modified into $\rho_0 = 1$, $\bm{u}_0 = 0$, $p_0 = 1$, and $\bm{B}_0 = 0.01\bm{e}_x\sin(2\pi z/L)$. The system was evolved for at least twelve large-eddy turnover times to reach the saturated phase of SSDs. Subsequently, the simulations were configured to output a sequence of consecutive data files containing both primitive variables and the driving force for the \textit{a posteriori} analysis of numerical dissipation.

\begin{figure*}[t!]
    \centering
    \includegraphics[width=\linewidth]{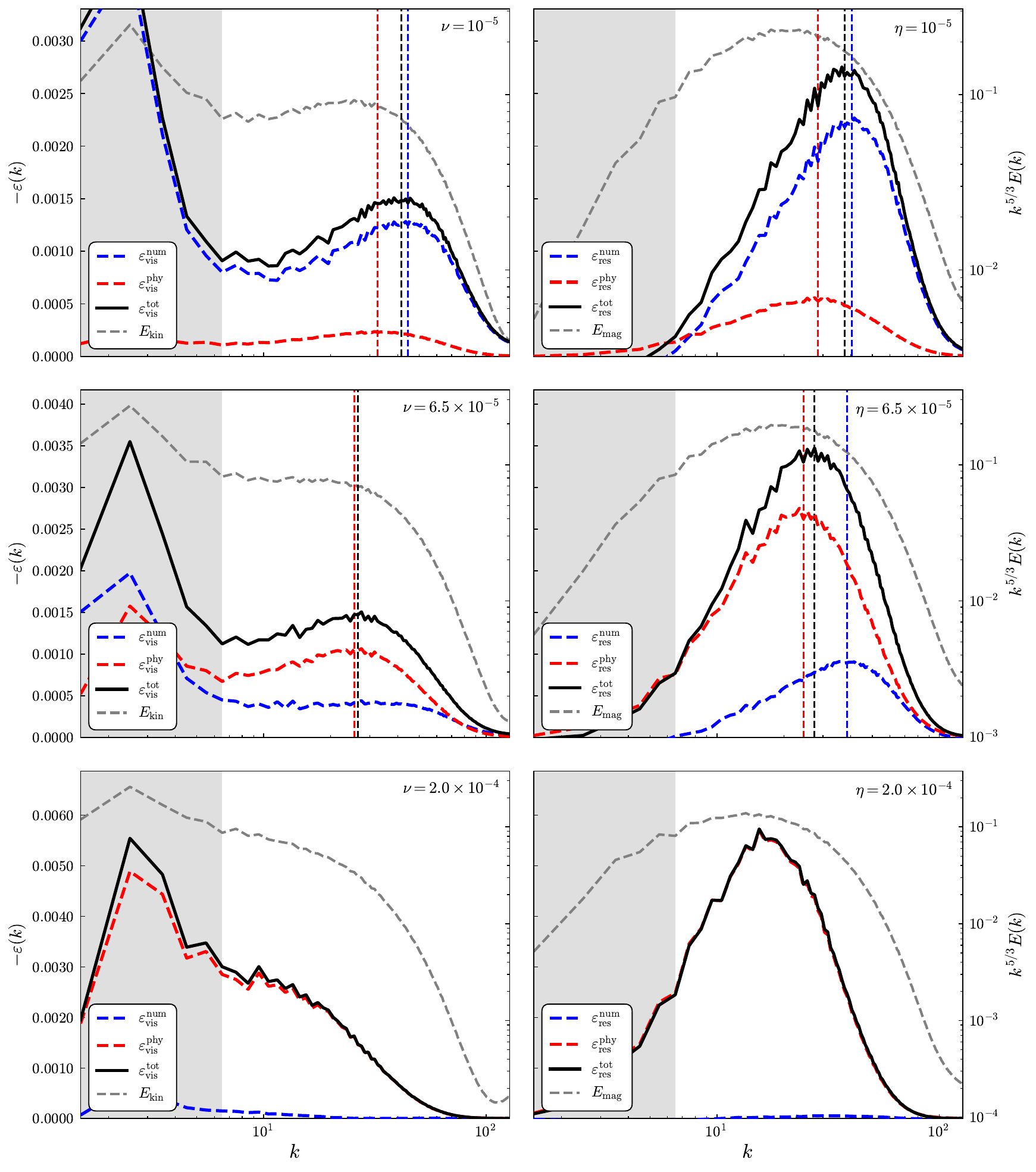}
    \caption{Shell-integrated dissipation spectra in $256^3$ simulations of the small-scale dynamo with $\mathrm{Pm}=1$ using PPM reconstruction. From top to bottom, the physical viscosity and resistivity were set to $\nu=\eta=10^{-5}$, $6.5\times10^{-5}$, and $2\times10^{-4}$, respectively. The left column shows the viscous dissipation spectra $\varepsilon_{\mathrm{vis}}(k)$ with the $k^{5/3}$-compensated kinetic energy spectra $E_{\mathrm{kin}}(k)$, whereas the right column shows the resistive dissipation spectra $\varepsilon_{\mathrm{res}}(k)$ with the $k^{5/3}$-compensated magnetic energy spectra $E_{\mathrm{mag}}(k)$. The red and blue dashed curves represent the physical and numerical dissipation spectra, respectively, while the black solid curves correspond to the total dissipation. The gray shaded background indicates the energy-injection scale. The peak dissipation wavenumber of the numerical dissipation, indicated by the vertical dashed line, remains nearly constant across simulations, corresponding to a reduced wavenumber of $\kappa\approx \pi/3$.}
    \label{fig:diss-spc-Pm1}
\end{figure*}

Figure~\ref{fig:diss-spc-Pm1} reveals a clear transition from numerically dominated to physically dominated dissipation as the physical dissipation coefficients $\nu=\eta$ increase. In theory, the kinetic energy spectra in turbulent SSDs are expected to follow Kolmogorov scaling, i.e., $E_{\mathrm{kin}}(k)\propto k^{-5/3}$ \citep{Beresnyak2012, Warnecke2023}. Accordingly, this figure shows the $k^{5/3}$-compensated energy spectra, for which $k^{5/3}E_{\mathrm{kin}}(k)=\mathrm{const}$ corresponds to the $-5/3$ Kolmogorov scaling and thus suggests the presence of an inertial range. In the low-dissipation case with $\nu=\eta=10^{-5}$, the kinetic energy spectrum exhibits approximate $-5/3$ scaling, accompanied by a pronounced bottleneck effect \citep{Falkovich1994, Lohse1995, Donzis2010}. Meanwhile, the dissipation is concentrated primarily at high wavenumbers, dissipating energy transferred by the turbulent cascade. The dissipation scales, defined as the peak wavenumbers of the total dissipation spectra $\varepsilon^{\mathrm{tot}}(\bm{k})=\varepsilon^{\mathrm{phy}}(\bm{k})+\varepsilon^{\mathrm{num}}(\bm{k})$, remain well separated from the energy-injection scale. Beyond these dissipation scales, both kinetic and magnetic energy are efficiently dissipated, as reflected in the rapid roll-off of the compensated energy spectra. In this case, the total dissipation spectra are dominated by numerical dissipation. Increasing the coefficients to $\nu=\eta=6.5\times10^{-5}$ produces a clear $-5/3$ scaling over $k \approx 7$-$25$. In this regime, the physical dissipation becomes larger than the numerical dissipation over the dissipation range. When the physical dissipation coefficients are further increased to $\nu=\eta=2\times10^{-4}$, the kinetic energy spectrum becomes noticeably steeper than $k^{-5/3}$, while the high-wavenumber viscous dissipation peak disappears. This indicates that the scale separation is no longer sufficient and that the inertial range has effectively disappeared.

It is also worth noting that the peak of the numerical dissipation spectrum remains nearly fixed at $k\approx 40$, corresponding to a reduced wavenumber of $\kappa\approx\pi/3$, independent of the physical dissipation coefficients. This behavior is consistent with the ADR analysis of the PPM scheme shown in Figure~\ref{fig:adr-b}, which suggests that significant numerical dissipation begins to emerge at this scale, indicating that numerical dissipation effectively introduces a resolution-dependent cutoff scale. Moreover, the physical dissipation spectrum peaks at smaller wavenumbers than the numerical dissipation spectrum, indicating that the two mechanisms dissipate energy primarily at different scales. Specifically, the characteristic scale of physical dissipation depends on the adopted dissipation coefficients, whereas numerical dissipation remains concentrated near the grid scale. Therefore, the effect of numerical dissipation cannot be reduced to a single effective dissipation coefficient or simply interpreted as an additional physical-like dissipation term.

\begin{figure*}[t!]
    \centering
    \includegraphics[width=\linewidth]{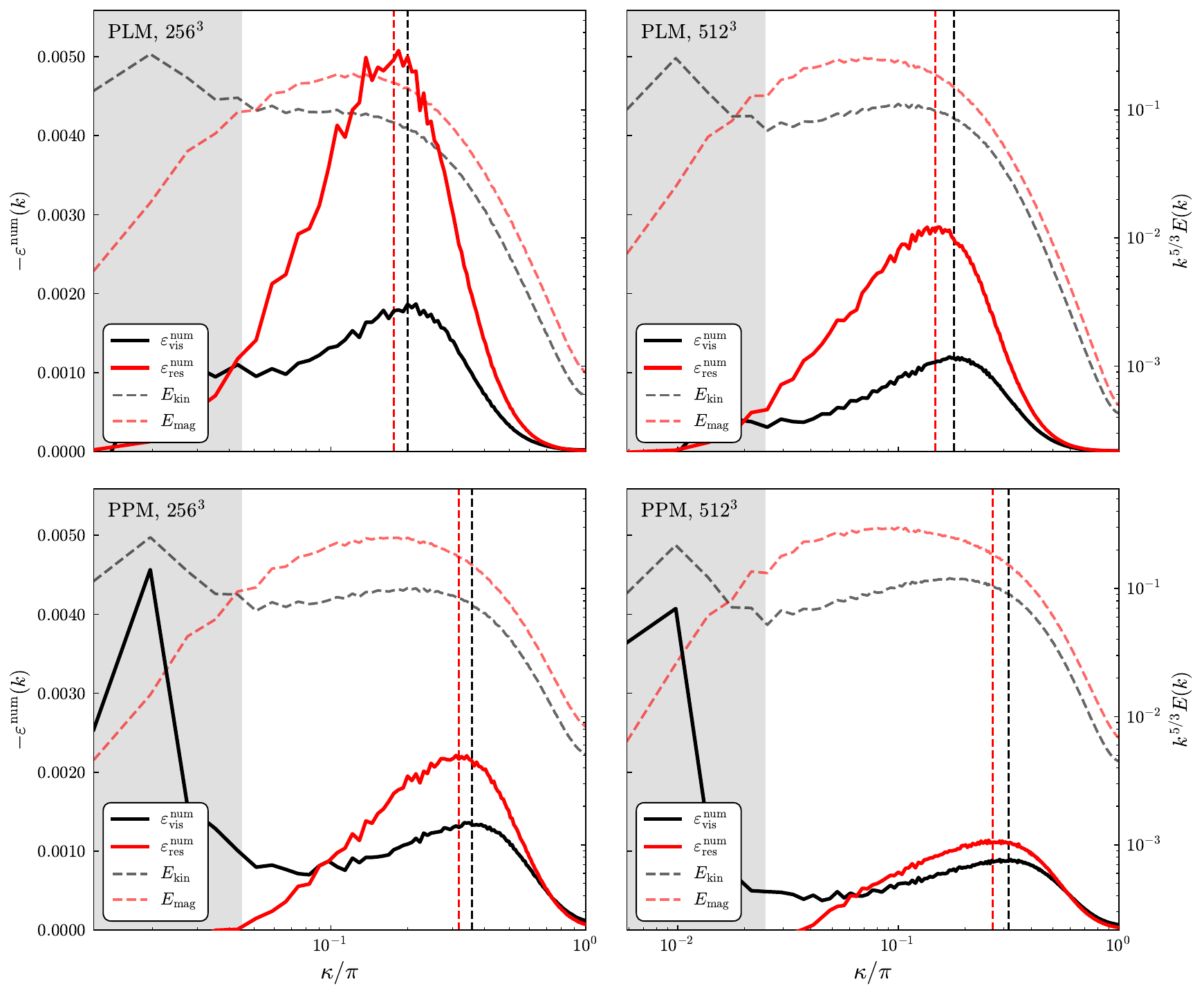}
    \caption{Shell-integrated numerical dissipation spectra and $k^{5/3}$-compensated energy spectra in ideal MHD simulations of the small-scale dynamo with PLM and PPM reconstructions. The top and bottom rows correspond to the PLM and PPM reconstructions, respectively, while the left and right columns show the $256^3$ and $512^3$ resolution runs. The solid black and red curves correspond to the numerical viscous and resistive dissipation spectra, $\varepsilon_{\mathrm{vis}}^{\mathrm{num}}(k)$ and $\varepsilon_{\mathrm{res}}^{\mathrm{num}}(k)$, respectively. The dashed curves represent the $k^{5/3}$-compensated kinetic and magnetic energy spectra. The horizontal axis shows the reduced wavenumber $\kappa = 2\pi kL/N$, with $\kappa=\pi$ corresponding to the Nyquist wavenumber of the numerical grid, while the gray shaded area indicates the energy-injection range. The peak wavenumber of the dissipation spectra, indicated by the vertical dashed lines, is located at a reduced wavenumber of approximately $\kappa \approx \pi/6$ for the PLM runs and $\kappa \approx \pi/3$ for the PPM runs.}
    \label{fig:diss-spc-ideal}
\end{figure*}

Figure~\ref{fig:diss-spc-ideal} shows that the numerical dissipation remains concentrated near the grid scale in the ideal MHD simulations (or ILES), with the peak dissipation wavenumber depending on the reconstruction scheme: the PLM runs peak at approximately $\kappa\approx\pi/6$, whereas the PPM runs peak at $\kappa\approx\pi/3$, beyond which the energy spectra roll off rapidly. This behavior is again consistent with the ADR analysis in Figure~\ref{fig:adr-b}, which indicates the onset of significant numerical dissipation at similar reduced wavenumbers for the two schemes, regardless of the grid resolution. In all cases, the numerical resistive dissipation is stronger than the numerical viscous dissipation and peaks at slightly larger scales. Specifically, increasing the resolution from $256^3$ to $512^3$ and replacing PLM with PPM both reduce the peak amplitude of the numerical dissipation spectra by roughly a factor of two. Moreover, this reduction is accompanied by a progressively stronger bottleneck effect in the compensated kinetic energy spectra. More specifically, the kinetic energy spectrum in the $256^3$ run with PLM reconstruction exhibits approximate $-5/3$ Kolmogorov scaling, whereas the remaining three simulations with weaker numerical dissipation all produce evident bottleneck effects. The bottleneck effect is most pronounced in the $512^3$ run with PPM reconstruction, where the numerical dissipation becomes too weak to terminate the turbulent cascade efficiently, leading to a noticeable departure from the expected $-5/3$ scaling.

These results highlight both the strengths and the limitations of ILES in simulations of MHD turbulence. On the one hand, Godunov-type methods with PLM reconstruction are widely used in high-resolution ILES of astrophysical MHD turbulence \citep{Grete2023, Beattie2025, Kempski2025}. This may be attributed to the relatively large numerical dissipation of PLM, which effectively acts as an implicit SGS stress that dissipates the energy transferred by the turbulent cascade. On the other hand, the fact that PLM's numerical dissipation peaks at $\kappa\approx\pi/6$ implies that the effective resolution for the inertial range is only about one-sixth of the grid resolution, indicating a substantial computational overhead. For reference, increasing the numerical resolution by a factor of six in a three-dimensional simulation corresponds to an increase in computational cost by roughly a factor of $6^4\approx10^3$. By contrast, PPM approximately doubles the effective resolution, whereas its numerical dissipation appears too weak for ILES. As a result, a more effective strategy for simulating MHD turbulence may be to adopt advanced numerical schemes with high spectral resolution, combined with explicit LES incorporating SGS models whose dissipation is concentrated at smaller scales.

\subsection{Anomalous Numerical Dissipation}\label{sec:anomalous}

In this section, we further decompose the energy and dissipation spectra into Cartesian components in order to characterize the anisotropic spectral properties of numerical dissipation and their impact on anisotropic turbulent energy spectra. We define the component-wise energy spectra as
\begin{equation}
    E_{\mathrm{kin},i}(\bm{k}) = \frac{1}{2}\left|\tilde{w}_i(\bm{k})\right|^2,\ E_{\mathrm{mag},i}(\bm{k}) = \frac{1}{2}\left|\tilde{B}_i(\bm{k})\right|^2,
\end{equation}
and the component-wise dissipation spectra as
\begin{equation}
    \begin{aligned}
    \mathscr{D}_{\mathrm{vis},i}(\bm{k}) & = \Re\left[\tilde{u}_i^*(\bm{k})\tilde{D}_{\mathrm{vis},i}(\bm{k})\right], \\
    \mathscr{D}_{\mathrm{res},i}(\bm{k}) & = \Re\left[\tilde{B}_i^*(\bm{k})\tilde{D}_{\mathrm{res},i}(\bm{k})\right],
    \end{aligned}
\end{equation}
such that the (total) energy and dissipation spectra can be written as
\begin{equation}
    E(\bm{k}) = \sum_{i=1}^{3} E_{i}(\bm{k}),\ \mathscr{D}(\bm{k}) = \sum_{i=1}^{3} \mathscr{D}_{i}(\bm{k}),
\end{equation}
respectively.

We focus here on \texttt{AthenaK} simulations of Alfv\'enic MHD turbulence in the presence of a strong background magnetic field. In contrast to \texttt{AthenaPK}, the background magnetic field in \texttt{AthenaK} is hardcoded to be aligned with the $z$-direction, i.e., $\bm{B}_0 = B_0 \bm{e}_z$. The resulting axisymmetry of the system about the background magnetic field naturally motivates a decomposition of energy and dissipation spectra into components perpendicular and parallel to $\bm{B}_0$. Specifically, the perpendicular spectra are defined as the sum of the $x$- and $y$-components,
\begin{equation}
    E_{\perp}(\bm{k}) = E_x(\bm{k}) + E_y(\bm{k}), \
    \mathscr{D}_{\perp}(\bm{k}) = \mathscr{D}_x(\bm{k}) + \mathscr{D}_y(\bm{k}),
\end{equation}
and the parallel spectra are defined as the $z$-component,
\begin{equation}
    E_{\parallel}(\bm{k}) = E_z(\bm{k}), \
    \mathscr{D}_{\parallel}(\bm{k}) = \mathscr{D}_z(\bm{k}).
\end{equation}
In contrast to the implementation of the driving force in simulations using \texttt{AthenaPK} described in Section~\ref{sec:anisotropic}, the \texttt{AthenaK} simulations considered here adopt a purely solenoidal and isotropic driving force with an energy injection rate of $\dot{\varepsilon}_{\mathrm{inj}}=0.1$ and a forcing spectrum proportional to $k^{-5/3}$ for $k_{\mathrm{low}} \le k \le k_{\mathrm{high}}$, with $k_{\mathrm{low}}=1.0$ and $k_{\mathrm{high}}=2.0$ (normalized to the box size $L=1.0$). The stochastic driving force evolves in time according to an OU process with an autocorrelation time of $T_{\mathrm{corr}}=0.5$. The simulations were initialized from a spatially uniform state with $\rho_0 = 1.0$, $\bm{u}_0 = 0$, and $\bm{B}_0 = B_0\bm{e}_z$, where the initial plasma beta was fixed at $\beta_0=2p_0/B_0^2=5.0$, corresponding to a strong background magnetic field. For the isothermal EoS, the constant sound speed is set to $c_{\mathrm{s}} = 10.0$, yielding the initial pressure $p_0 = c_{\mathrm{s}}^2 \rho_0 = 100.0$; for the adiabatic EoS with $\gamma = 5/3$, the initial pressure is instead set to $p_0 = 60.0$, yielding an identical sound speed of $c_{\mathrm{s}}=\sqrt{\gamma p_0/\rho_0}=10.0$. Each simulation was evolved for at least ten large-eddy turnover times to reach a statistically stationary state, after which a sequence of consecutive data files was recorded for the \textit{a posteriori} analysis of numerical dissipation.

\begin{figure*}[t!]
    \centering
    \includegraphics[width=\linewidth]{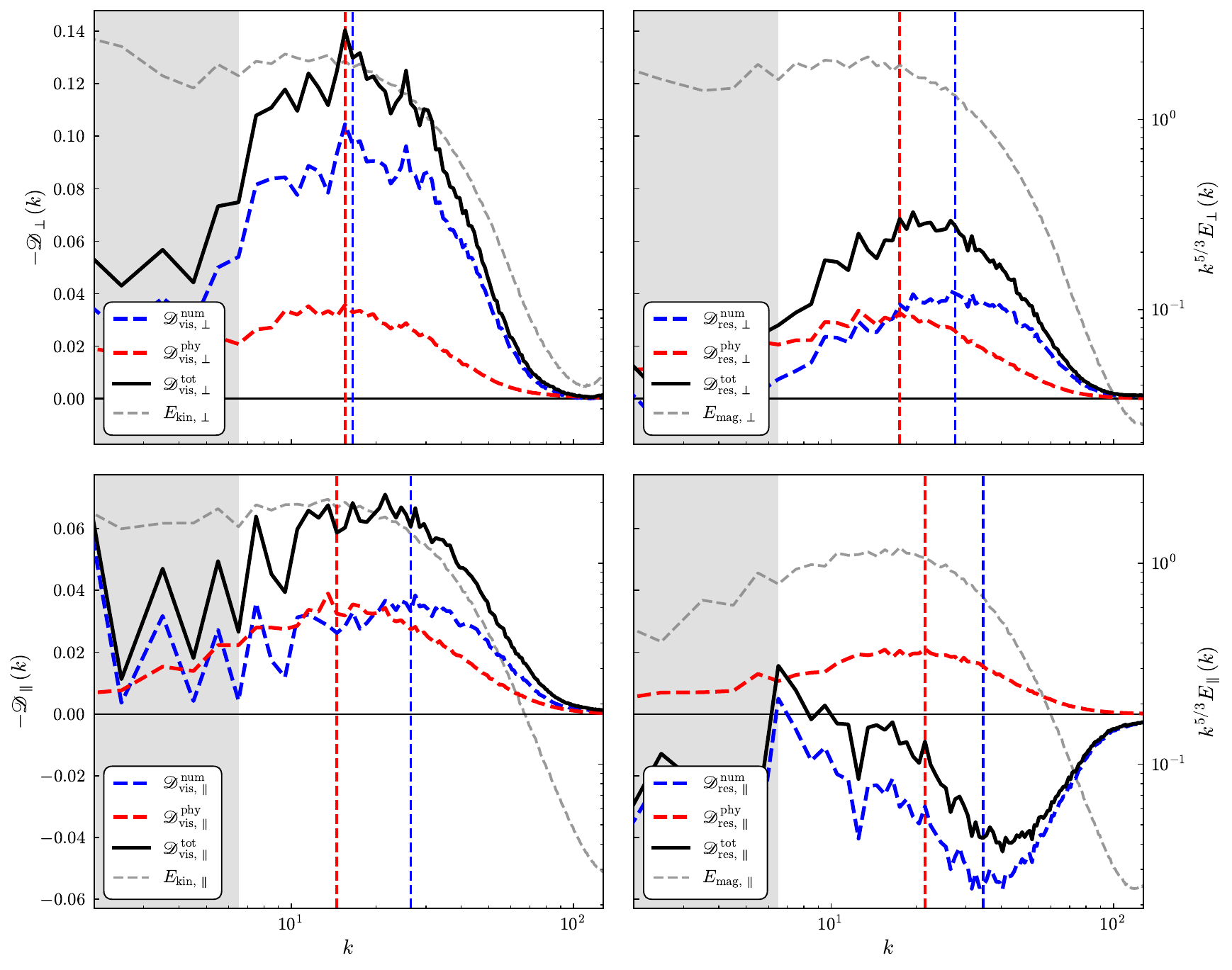}
    \caption{Shell-integrated component-wise dissipation spectra and $k^{5/3}$-compensated energy spectra in a $256^3$ simulation of Alfv\'enic MHD turbulence with $\nu=\eta=10^{-4}$, an isothermal EoS, PPM reconstruction, and the HLLD Riemann solver. The top row shows the spectra perpendicular to the background magnetic field, while the bottom row shows the parallel spectra. The left column presents viscous dissipation spectra together with kinetic energy spectra, and the right column presents resistive dissipation spectra together with magnetic energy spectra. Blue dashed curves denote numerical dissipation, red dashed curves physical dissipation, black solid curves total dissipation, and gray dashed curves the corresponding energy spectra. The gray shaded regions indicate the energy-injection scales. A notable feature is that the numerical resistive dissipation spectrum for the parallel component exhibits pronounced anomalous numerical dissipation, i.e., $\mathscr{D}^{\mathrm{num}}_{\mathrm{res},\parallel}(k)>0$ over the entire wavenumber range, implying spurious numerical injection of magnetic energy rather than dissipation.}
    \label{fig:anomalous}
\end{figure*}

Figure~\ref{fig:anomalous} shows the shell-integrated component-wise dissipation spectra and the corresponding $k^{5/3}$-compensated energy spectra for the \texttt{AthenaK} simulation of Alfv\'enic MHD turbulence with an isothermal EoS, PPM reconstruction, and the HLLD Riemann solver. For all components, the numerical dissipation spectra generally peak at higher wavenumbers than their physical counterparts, consistent with the behavior found in Section~\ref{sec:spectral-analysis}. In addition, a particularly striking feature appears in the component parallel to the background magnetic field, where the numerical resistive dissipation exhibits anomalous anti-dissipative behavior: instead of dissipating magnetic energy like its physical counterpart with $\mathscr{D}_{\mathrm{res}}(k)<0$, the parallel numerical resistive spectrum remains positive, with $\mathscr{D}^{\mathrm{num}}_{\mathrm{res},\parallel}(k)>0$ over almost the entire wavenumber range, indicating that the truncation error acts as a spurious magnetic energy injection, especially at high wavenumbers. For the relatively small physical dissipation coefficients with $\nu=\eta=10^{-4}$ adopted here, this anomalous numerical dissipation dominates the total parallel resistive dissipation. Consequently, while the other component-wise energy spectra remain close to the $-5/3$ Kolmogorov scaling over the inertial range, the parallel magnetic energy spectrum is noticeably shallower. This slightly shallower slope is possibly attributable to the spurious energy injection produced by the anomalous numerical dissipation at intermediate to high wavenumbers.

\begin{figure*}[t!]
    \centering
    \includegraphics[width=\linewidth]{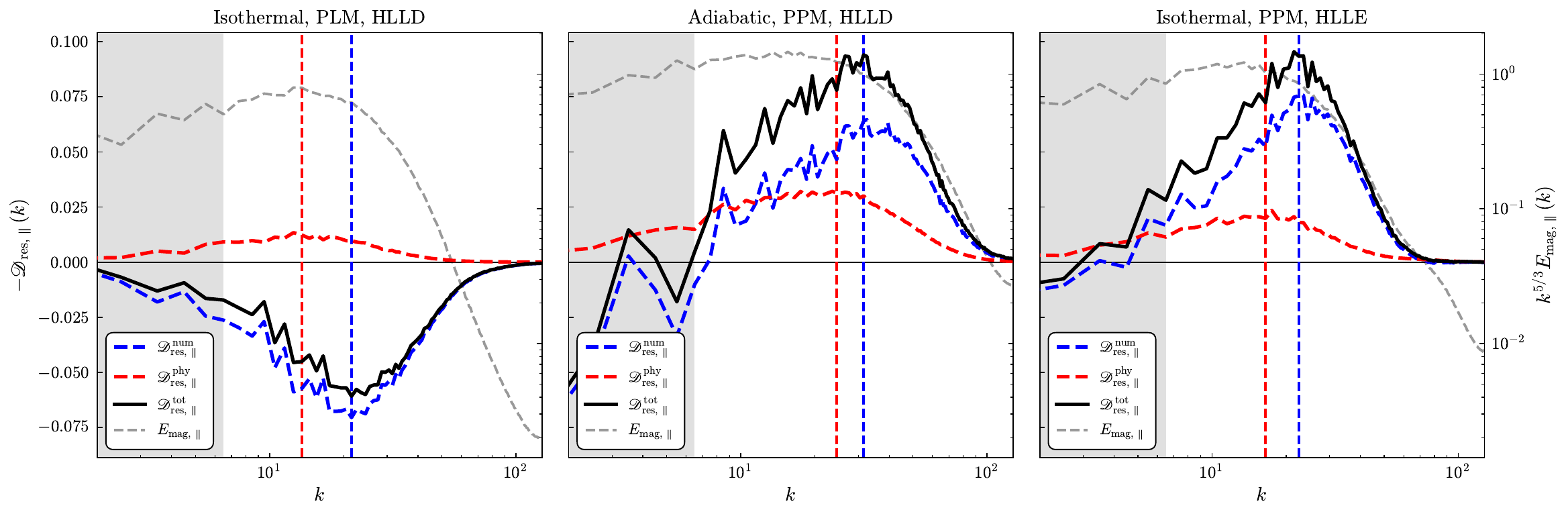}
    \caption{Shell-integrated parallel resistive spectra $\mathscr{D}_{\mathrm{res},\parallel}(k)$ and $k^{5/3}$-compensated magnetic energy spectra $E_{\mathrm{mag},\parallel}(k)$ in $256^3$ simulations of Alfv\'enic MHD turbulence with $\nu=\eta=10^{-4}$ and: an isothermal EoS, PLM reconstruction, and HLLD Riemann solver (left); an adiabatic EoS, PPM reconstruction, and HLLD Riemann solver (middle); an isothermal EoS, PPM reconstruction, and HLLE Riemann solver (right). The curves and gray shaded regions have the same meanings as in Figure~\ref{fig:anomalous}. Compared with Figure~\ref{fig:anomalous}, the anomalous behavior of the parallel numerical resistive spectrum remains in the isothermal HLLD case even with a different reconstruction scheme, but is not observed in the adiabatic HLLD and isothermal HLLE cases. This result suggests that the anomalous numerical dissipation might be associated with the HLLD Riemann solver for isothermal MHD equations \citep{Mignone2007HLLD}.}
    \label{fig:anomalous-multi}
\end{figure*}

Furthermore, Figure~\ref{fig:anomalous-multi} compares the shell-integrated parallel resistive spectra among three similar \texttt{AthenaK} simulations of Alfv\'enic MHD turbulence with: an isothermal EoS, PLM reconstruction, and HLLD Riemann solver (left); an adiabatic EoS, PPM reconstruction, and HLLD Riemann solver (middle); and an isothermal EoS, PPM reconstruction, and HLLE Riemann solver (right). The comparison shows that the anomalous anti-dissipative behavior of the parallel numerical resistive spectrum is still observed with PLM reconstruction, while it disappears when either the EoS is changed from isothermal to adiabatic or the Riemann solver is changed from HLLD to HLLE. This indicates that the anomalous numerical dissipation is most likely associated with the combination of an isothermal EoS and HLLD Riemann solver.

It is worth noting that the HLLD implementations for the adiabatic and isothermal EoSs in \texttt{AthenaK} are different, corresponding to \citet{Miyoshi2005} and \citet{Mignone2007HLLD}, respectively, whereas the HLLE Riemann solver does not depend on the choice of EoS \citep{Einfeldt1988, Einfeldt1991}. This suggests that the anomalous behavior may be specifically related to the isothermal HLLD Riemann solver of \citet{Mignone2007HLLD}, although the precise mechanism requires further investigation. In addition, the simulation showing anomalous numerical dissipation also exhibits a slightly shallower parallel magnetic energy spectrum than the other two simulations, suggesting that the spurious numerical energy injection may indeed affect the turbulent cascade. Despite the anti-dissipative contribution in the parallel resistive spectrum, the net numerical energy dissipation spectrum $\varepsilon^{\mathrm{num}}_{\mathrm{tot}}(k) = \varepsilon^{\mathrm{num}}_{\mathrm{vis}}(k) + \varepsilon^{\mathrm{num}}_{\mathrm{res}}(k)$ remains dissipative, thereby maintaining the overall numerical stability of the simulation. The identification of anomalous numerical dissipation further demonstrates the ability of the proposed framework to comprehensively characterize the properties of numerical dissipation. A similar indication of negative numerical resistivity was also reported by \citet{Rembiasz2017}, albeit in simulations of linear waves with an adiabatic EoS. More generally, the presence of such anti-dissipative behavior suggests potential limitations on the applicability of ILES under these circumstances, in which the anomalous numerical dissipation cannot be consistently interpreted as either an additional physical dissipation term or an implicit SGS model.

\section{High-fidelity DNS} \label{sec:dns}

The previous section analyzed the statistical properties of numerical dissipation in MHD turbulence simulations and concluded that numerical dissipation cannot, in general, be interpreted as an effective physical dissipation. It also demonstrated that numerical dissipation may become highly anisotropic in anisotropic MHD turbulence. Therefore, achieving high-fidelity DNS requires minimizing the impact of numerical dissipation on the turbulence. A practical criterion is that physical dissipation should dominate over numerical dissipation across the entire spectrum. In this section, turbulent MRI-driven dynamos are used as a representative example of anisotropic MHD turbulence to investigate how this criterion can be met in practice, thereby providing practical guidance for achieving high-fidelity DNS.

\begin{figure*}[t!]
    \centering
    \includegraphics[width=\linewidth]{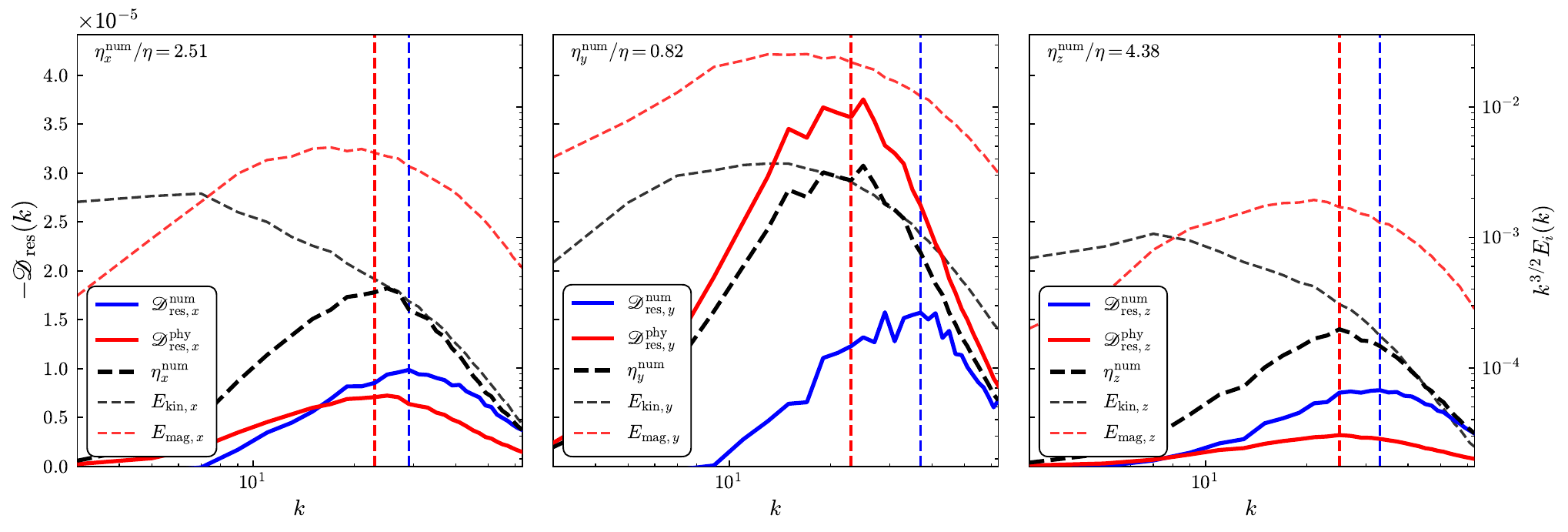}
    \caption{Shell-integrated component-wise resistive dissipation spectra and $k^{3/2}$-compensated energy spectra in a shearing-box simulation of an MRI-driven dynamo with $\mathrm{Re}=10^4$, $\mathrm{Pm}=20$, and PPM reconstruction. The left, middle, and right panels correspond to the $x$-, $y$-, and $z$-components, respectively. In each panel, the blue and red solid curves show the numerical and physical resistive dissipation spectra, $\mathscr{D}^{\mathrm{num}}_{\mathrm{res},i}(k)$ and $\mathscr{D}^{\mathrm{phy}}_{\mathrm{res},i}(k)$, respectively, while the black dashed curve shows $\eta_i^{\mathrm{num}}\hat{\mathscr{D}}^{\mathrm{phy}}_{\mathrm{res},i}(k)$, i.e., the physical resistive dissipation spectrum scaled by $\eta_i^{\mathrm{num}}/\eta$, where $\eta_i^{\mathrm{num}}$ is defined in Equation~\eqref{eq:eta-num}. The gray and red dashed curves denote the component-wise kinetic and magnetic energy spectra, $k^{3/2}E_{\mathrm{kin},i}(k)$ and $k^{3/2}E_{\mathrm{mag},i}(k)$, respectively. The vertical blue and red dashed lines mark the peak wavenumbers of the numerical and physical dissipation spectra, respectively.}
    \label{fig:fit}
\end{figure*}

Figure~\ref{fig:fit} compares the component-wise physical and numerical resistive dissipation spectra in a $256\times 256\times 128$ \texttt{AthenaK} simulation of the turbulent MRI-driven dynamo with $\mathrm{Re}=10^4$, $\mathrm{Pm}=20$, and PPM reconstruction. In this regime, numerical viscous dissipation is negligible compared to its physical counterpart; therefore, the discussion below is focused on resistive dissipation. The component-wise compensated kinetic and magnetic energy spectra show pronounced anisotropy, with the $y$-component of magnetic energy exceeding the $x$- and $z$-components by roughly an order of magnitude across the entire spectrum. This behavior is a consequence of the background shear flow in the shearing-box simulation, which preferentially amplifies the azimuthal ($y$-component) magnetic field. A similar anisotropy is observed in the physical resistive dissipation spectra, with $\mathscr{D}^{\mathrm{phy}}_{\mathrm{res},y}(k)$ substantially exceeding its $x$- and $z$-counterparts. The numerical resistive dissipation spectra, by contrast, exhibit much weaker variation among the three components. As a result, physical resistive dissipation clearly dominates over its numerical counterpart in the $y$-component, whereas in the $x$- and $z$-components the numerical dissipation remains comparable to, or even exceeds, the physical dissipation over most of the dissipation range.

Another notable feature is that the numerical and physical dissipation spectra exhibit distinctly different spectral profiles. In all three components, the numerical dissipation peaks at larger wavenumbers than the physical dissipation, indicating that it is concentrated at smaller scales, similar to the behavior observed in the simulations of SSDs discussed in Section~\ref{sec:spectral-analysis}. This indicates that numerical dissipation cannot be represented by a single effective resistivity coefficient obtained from a direct fit to the physical dissipation spectrum. To quantify the numerical resistivity relative to its physical counterpart, we introduce a coefficient-free physical resistive dissipation spectrum based on the component-wise dissipation spectra defined in Section~\ref{sec:anomalous} as
\begin{equation}
\hat{\mathscr{D}}^{\mathrm{phy}}_{\mathrm{res},i}(\bm{k})
=
\Re\left[\mathscr{F}(B_i)^*\mathscr{F}(\nabla^2 B_i)\right],
\end{equation}
such that the physical resistive dissipation spectrum can be written as
\begin{equation}
\mathscr{D}^{\mathrm{phy}}_{\mathrm{res},i}(\bm{k})
=
\eta \hat{\mathscr{D}}^{\mathrm{phy}}_{\mathrm{res},i}(\bm{k}).
\end{equation}
Using the shell-integrated spectra $\mathscr{D}^{\mathrm{num}}_{\mathrm{res},i}(k)$ and $\hat{\mathscr{D}}^{\mathrm{phy}}_{\mathrm{res},i}(k)$, we subsequently define an upper bound on the effective numerical resistivity in each component, denoted by $\eta_i^{\mathrm{num}}$, as the smallest coefficient satisfying
\begin{equation}
\forall k, \eta_i^{\mathrm{num}}\left|\hat{\mathscr{D}}^{\mathrm{phy}}_{\mathrm{res},i}(k)\right|
\ge
\left|\mathscr{D}^{\mathrm{num}}_{\mathrm{res},i}(k)\right|,
\end{equation}
or equivalently,
\begin{equation}\label{eq:eta-num}
\eta_i^{\mathrm{num}}
=
\max_k
\left|
\frac{\mathscr{D}^{\mathrm{num}}_{\mathrm{res},i}(k)}
{\hat{\mathscr{D}}^{\mathrm{phy}}_{\mathrm{res},i}(k)}
\right|.
\end{equation}
In other words, $\eta_i^{\mathrm{num}}$ is the smallest coefficient such that the scaled physical dissipation $\eta_i^{\mathrm{num}}\hat{\mathscr{D}}^{\mathrm{phy}}_{\mathrm{res},i}(k)$ dominates over the numerical dissipation $\mathscr{D}^{\mathrm{num}}_{\mathrm{res},i}(k)$ over the entire spectrum, as represented by the black dashed curve in Figure~\ref{fig:fit}.

It is then convenient to normalize $\eta_i^{\mathrm{num}}$ by the physical resistivity $\eta$ and introduce the dimensionless component-wise upper bound
\begin{equation}
\xi_i = \frac{\eta_i^{\mathrm{num}}}{\eta} =
\max_k
\left|
\frac{\mathscr{D}^{\mathrm{num}}_{\mathrm{res},i}(k)}
{\mathscr{D}^{\mathrm{phy}}_{\mathrm{res},i}(k)}
\right|,
\end{equation}
which gives the smallest factor by which the physical resistive dissipation spectrum must be scaled to dominate the numerical dissipation spectrum at all wavenumbers. Thus, $\xi_i<1$ indicates that, in the $i$-th component, physical resistive dissipation dominates numerical dissipation at all wavenumbers, whereas $\xi_i\ge 1$ implies that numerical dissipation cannot be neglected relative to physical dissipation over at least part of the spectrum. The values of $\xi_i=\eta_i^{\mathrm{num}}/\eta$ are shown in the upper-left corner of each panel in Figure~\ref{fig:fit}, which demonstrate a clear anisotropy in the relative importance of numerical dissipation with $\xi_x=2.51$, $\xi_y=0.82$, and $\xi_z=4.38$. Among the three components, only the $y$-component satisfies $\xi_i<1$, showing that the $y$-component resistive dissipation remains physically dominated across the entire spectrum. By contrast, both the $x$- and $z$-components have $\xi_i>1$, most notably the $z$-component. Therefore, in this simulation, numerical dissipation cannot be neglected compared to physical dissipation.

\begin{figure}[t!]
    \centering
    \includegraphics[width=\linewidth]{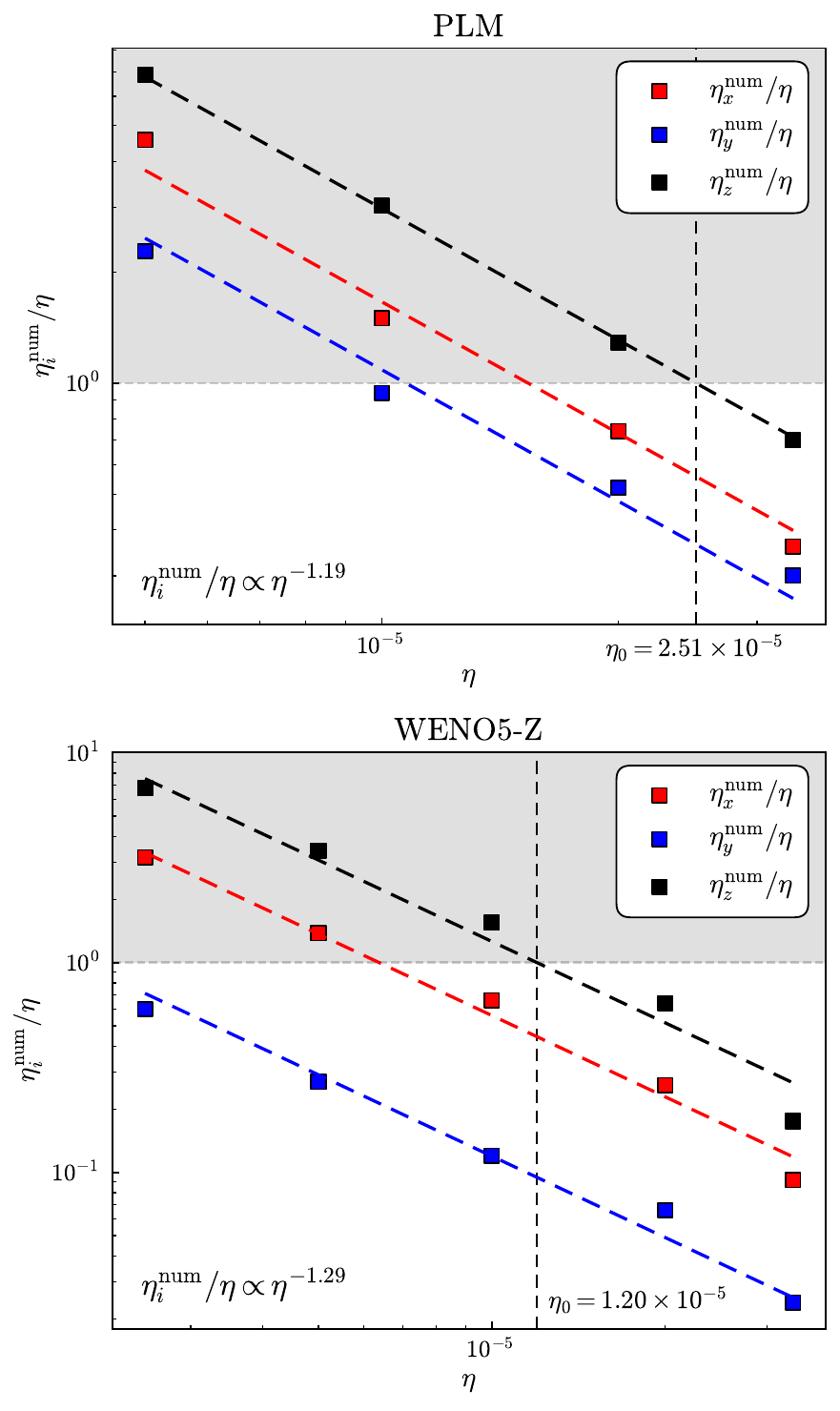}
    \caption{Scaling laws of the dimensionless component-wise upper bound $\xi_i=\eta_i^{\mathrm{num}}/\eta$ with respect to physical resistivity $\eta$ in $256\times 256\times 128$ simulations of MRI-driven dynamo at $\mathrm{Re}=10^4$. The top and bottom panels correspond to the PLM and WENO5-Z reconstruction schemes, respectively. The red, blue, and black symbols represent $\xi_x$, $\xi_y$, and $\xi_z$, while the dashed lines indicate the corresponding power-law fits, yielding $\xi_i \propto \eta^{-1.2}$ and $\xi_i \propto \eta^{-1.3}$ for PLM and WENO5-Z, respectively. The gray shaded region indicates $\xi_i>1$, where numerical dissipation is non-negligible compared to physical dissipation over at least part of the spectrum. The vertical dashed line indicates the threshold $\eta_0$ required for high-fidelity DNS, defined by $\max_i(\xi_i)=1$. This yields $\eta_0=2.51\times10^{-5}$ for PLM and $\eta_0=1.20\times10^{-5}$ for WENO5-Z, corresponding to magnetic Reynolds numbers of $3.98\times10^4$ and $8.33\times10^4$, respectively.}
    \label{fig:scaling}
\end{figure}

Figure~\ref{fig:scaling} shows that the relative numerical resistivity follows an approximate power-law dependence on the physical resistivity, with $\xi_i \propto \eta^{-1.2}$ for PLM and $\xi_i \propto \eta^{-1.3}$ for WENO5-Z. This indicates a negative correlation between numerical and physical resistivity, consistent with the results in Section~\ref{sec:spectral-analysis}. At fixed resolution and $\eta$, the WENO5-Z reconstruction yields systematically smaller $\xi_i$ than PLM in all components, confirming that the higher-order scheme indeed introduces weaker numerical dissipation. Consequently, the threshold for achieving high-fidelity DNS, i.e., the resistivity at which physical dissipation dominates numerical dissipation at all wavenumbers, is lower for WENO5-Z, with $\eta_0=2.51\times10^{-5}$ for PLM and $\eta_0=1.20\times10^{-5}$ for WENO5-Z, corresponding to $\mathrm{Rm}\approx 3.98\times10^4$ and $8.33\times10^4$, respectively. Therefore, higher-order schemes are essential for extending high-fidelity DNS to larger (magnetic) Reynolds numbers at fixed resolution.

\section{Conclusions} \label{sec:conclusions}

In this work, we have developed an \textit{a posteriori} framework for the comprehensive characterization of numerical dissipation in simulations of MHD turbulence. Implemented in the open-source Python package \texttt{PyMHD}, this framework directly estimates numerical viscosity and resistivity from simulation output data without invoking any \textit{a priori} empirical ansatz, and is therefore applicable to various numerical schemes and flow configurations. Within this residual-based framework, we proposed a novel targeted compact scheme (TCS) with a multi-stencil discontinuity detector (MSDD), providing accurate residual evaluation for simulations of compressible turbulence containing discontinuities with robust shock-capturing capability. By applying this framework to simulations of selected astrophysical MHD turbulence, we have, to the best of our knowledge, presented the first systematic characterization of the anisotropy and spectral properties of numerical dissipation.

The most important conclusion of this paper is that numerical dissipation in simulations of MHD turbulence cannot, in general, be interpreted as an additional physical-like dissipation with effective coefficients. Our analysis shows that the spectral profile of numerical dissipation is fundamentally different from those of physical viscosity and resistivity: its dissipation spectra generally peak at higher wavenumbers and remain concentrated closer to the grid scale. We also find that, in anisotropic turbulence, numerical dissipation itself becomes anisotropic, thereby introducing additional complexity into the interpretation of the simulation results. Moreover, the spectral properties of numerical dissipation are highly sensitive to the choice of reconstruction scheme, with PPM exhibiting an effective spectral resolution approximately twice that of PLM. Finally, under specific conditions, numerical dissipation can even exhibit anomalous anti-dissipative behavior that is possibly associated with the isothermal HLLD Riemann solver.

These results motivate a fundamental shift from traditional descriptions of numerical dissipation using effective coefficients to a more comprehensive characterization with, for instance, component-wise dissipation spectra. This provides a theoretical basis for interpreting MHD turbulence simulations by explicitly quantifying numerical dissipation, thereby distinguishing genuine physics from numerical artifacts. This framework also provides practical guidance for setting simulation parameters to achieve high-fidelity DNS. In addition, it establishes a robust methodology to evaluate the applicability and limitations of ILES in MHD turbulence, including its effective resolution for the inertial range, and whether the numerical dissipation is sufficient to serve as a reliable implicit SGS model. Furthermore, the proposed framework may facilitate the design of numerical schemes tailored for ILES in MHD turbulence to achieve controlled numerical dissipation with optimized spectral properties.

While the framework developed here is generally applicable to simulations of MHD turbulence, the current post-processing pipeline in \texttt{PyMHD} remains restricted to homogeneous turbulence and is not yet designed for strongly inhomogeneous systems such as stratified MRI-driven turbulence. In addition, the present implementation is not compatible with adaptive mesh refinement (AMR), which would require more sophisticated numerical treatment to handle nonuniform grids and coarse-fine interfaces in the residual evaluation. Finally, the MHD turbulence examined in this work spans only a limited subset of parameter space; extending the characterization to other regimes, such as SSDs at high magnetic Prandtl numbers, would require further investigation. These limitations also motivate several directions for future work, including applying the framework to a broader range of simulation configurations and numerical schemes to identify specific algorithms that are especially well suited for high-fidelity DNS of MHD turbulence; generalizing the current implementation to inhomogeneous turbulence and AMR; and conducting a more detailed investigation of the mechanisms underlying anomalous numerical dissipation through additional numerical experiments.

\begin{acknowledgments}
    This work is supported by the Science Challenge Project, No. TZ2025012; the National Key R\&D Program of China, Nos. 2022YFA1603200, 2022YFA1603201, and 2022YFA1603203; the National Natural Science Foundation of China, grants Nos. 12135001, 11921006, 11825502, and 12305267; the Strategic Priority Research Program of the Chinese Academy of Sciences, grant No. XDA25050900. B.Q. acknowledges support from the National Natural Science Funds for Distinguished Young Scholars, grant No. 11825502. The authors acknowledge the use of OpenAI's GPT-5.5 \citep{openai2026gpt55} and Anthropic's Claude Opus 4.8 \citep{anthropic2026opus48} for assistance in improving the clarity and readability of the text and proofreading during the preparation of this manuscript. All LLM-generated text was carefully reviewed and edited by the authors.
\end{acknowledgments}

\software{
    \texttt{PyMHD} \citep{Hua2026}, \texttt{AthenaK} \citep{Stone2026}, \texttt{AthenaPK} \citep{Grete2021, Holmen2024}, \texttt{yt} \citep{Turk2010}, \texttt{NumPy} \citep{Harris2020}, \texttt{JAX} \citep{jax2018}, \texttt{KDEpy} \citep{kdepy}
}


\appendix

\section{Godunov-type Methods}\label{appendix:basic-framework}

Built upon the finite volume method (FVM), the Godunov-type methods enforce conservation laws to machine precision by evolving cell-averaged variables and calculating interface fluxes via spatial reconstruction and (approximate) Riemann solvers. The general form of a system of conservation laws is given by \citep{Godlewski2021}
\begin{equation}\label{eq:conservation-law}
\frac{\partial u}{\partial t} + \nabla \cdot \mathbb{F} (u) = 0,
\end{equation}
where $u$ denotes the vector of conserved variables and $\mathbb{F}$ the flux functions. For a computational domain $\Omega$, let $\Omega_{i}$ denote a single cell and $t^{(n)}$ the $n$-th time step.

Godunov-type methods evolve the cell-averaged variables $\overline{u}_{i}$ using the integral form of the conservation laws with the semi-discrete form obtained by summing the numerical fluxes over all neighboring cells $\mathcal{N}(i)$:
\begin{equation}\label{eq:semi-discrete}
\frac{\mathrm{d} \overline{u}_{i}}{\mathrm{d} t} = - \frac{1}{|\Omega_i|} \sum_{j \in \mathcal{N}(i)} |\Gamma_{ij}| F^{\mathrm{num}}(u_{ij}, u_{ji}, \bm{n}_{ij}).
\end{equation}
Here, the numerical flux $F^{\mathrm{num}}(u_{ij}, u_{ji}, \bm{n}_{ij})$ approximates the physical flux across the interface $\Gamma_{ij}$ between cell $\Omega_i$ and a neighboring cell $\Omega_j$ with unit normal vector $\bm{n}_{ij}$, where $u_{ij}, u_{ji}$ are the (reconstructed) states on the inner and outer sides of the interface $\Gamma_{ij}$, respectively. Modern Godunov-type methods typically employ high-order interface reconstructions, such as the second-order piecewise linear method (PLM, \citealt{vanLeer1979}) and the third-order piecewise parabolic method (PPM, \citealt{Colella1984}). The numerical flux is often computed with efficient approximate Riemann solvers like Roe, HLLC, HLLE, or HLLD \citep{Toro2009}.

Equation~\eqref{eq:semi-discrete} represents a system of ordinary differential equations with respect to time, and can be rewritten compactly as $\mathrm{d}U/\mathrm{d}t = \mathcal{R}(U)$, where $U=\{\overline{u}_i\}$ denotes the global vector of conserved variables across all the cells and $\mathcal{R}(U)$ corresponds to the right-hand side of Equation~\eqref{eq:semi-discrete}. Explicit time integration is often employed to advance the solution from time $t^{(n)}$ to $t^{(n+1)}$ with time step $\Delta t$. The most straightforward time integrator is the forward Euler method
\begin{equation}
U^{(n+1)} = U^{(n)} + \Delta t \mathcal{R}(U^{(n)})
\end{equation}
with first-order accuracy. However, astrophysical MHD codes typically employ high-order time integrators, such as the second-order van Leer integrator and the family of strong stability preserving Runge-Kutta (SSP-RK) schemes \citep{Gottlieb2001}.

Finally, the time step is strictly constrained by the Courant-Friedrichs-Lewy (CFL) condition:
\begin{equation}\label{eq:cfl}
\Delta t = \mathrm{CFL}\cdot \min_{i} \left[ \sum_{k=1}^{d} \frac{\lambda_{\max, i}^{(k)}}{\Delta_i^{(k)}} \right]^{-1},
\end{equation}
where $\Delta_i^{(k)}$ and $\lambda_{\max, i}^{(k)}$ denote the cell width and the maximum signal speed along the $k$-th direction in cell $\Omega_i$, respectively, and $\mathrm{CFL}$ is the user-specified CFL number.

\section{Shock-capturing Reconstruction Schemes}\label{appendix:reconstruction}

PLM and PPM schemes typically rely on specific limiters to suppress spurious oscillations in the vicinity of discontinuities. However, application of such limiters induces a ``clipping'' effect at smooth local extrema, leading to accuracy degeneration and excessive numerical dissipation.

\subsection{The WENO-Z Scheme}

In contrast to PLM and PPM, high-order shock-capturing schemes, most notably the weighted essentially non-oscillatory (WENO) families \citep{Jiang1996, Balsara2000, Borges2008, Shu2009}, employ adaptive strategies to capture discontinuities while maintaining high-order accuracy in smooth regions. The classical fifth-order WENO-Z scheme \citep{Borges2008} utilizes a five-point stencil
\begin{equation}
\mathrm{S}^{(5)}_i=\{x_{i-2}, x_{i-1}, x_{i}, x_{i+1}, x_{i+2}\}
\end{equation}
to reconstruct the cell-interface values at (for instance) $x_{i+1/2}$. The shock-capturing capability is achieved via a convex combination of candidate reconstruction polynomials constructed on three underlying substencils:
\begin{equation}
\mathrm{S}^{(3,0)}_i=\{x_{i-2}, x_{i-1}, x_{i}\}, \
\mathrm{S}^{(3,1)}_i=\{x_{i-1}, x_{i}, x_{i+1}\}, \
\mathrm{S}^{(3,2)}_i=\{x_{i}, x_{i+1}, x_{i+2}\},
\end{equation}
as illustrated in Figure~\ref{fig:weno5}. Specifically, the reconstructed interface $u_{i+1/2}$ can be written as a weighted sum of three candidates with third-order accuracy (for brevity, the subscripts $i+1/2$ and $i$ on $w^{(k)}$, stencils, and associated quantities are dropped hereinafter):
\begin{equation}
u_{i+1/2} = w^{(0)} u_{i+1/2}^{(0)} + w^{(1)} u_{i+1/2}^{(1)} + w^{(2)} u_{i+1/2}^{(2)},
\end{equation}
where the normalized weights satisfy $w^{(0)}+w^{(1)}+w^{(2)} =1$ and $w^{(k)}\ge 0$. The candidate reconstruction polynomials from each substencil are given by:
\begin{equation}
u_{i+1/2}^{(0)} = \frac{1}{3} \overline{u}_{i-2} - \frac{7}{6} \overline{u}_{i-1} +\frac{11}{6} \overline{u}_{i},\
u_{i+1/2}^{(1)} = -\frac{1}{6} \overline{u}_{i-1} + \frac{5}{6} \overline{u}_{i} +\frac{1}{3} \overline{u}_{i+1},\
u_{i+1/2}^{(2)} = \frac{1}{3} \overline{u}_{i} + \frac{5}{6} \overline{u}_{i+1} -\frac{1}{6} \overline{u}_{i+2}.
\end{equation}

In smooth regions, the nonlinear weights $w^{(k)}$ are designed to converge to the linear weights $d^{(k)}$ with
\begin{equation}
d^{(0)} = \frac{1}{10}, d^{(1)} = \frac{3}{5}, d^{(2)} = \frac{3}{10}
\end{equation}
to recover the fifth-order accuracy. In the presence of discontinuities within the stencil $\mathrm{S}^{(5)}$, the nonlinear weights $w^{(k)}$ should be adapted according to the local smoothness of the solution on each substencil. This is quantified by the smoothness indicator $\beta^{(3,k)}$, originally proposed by \citet{Jiang1996}. These smoothness indicators\footnote{The definition and explicit forms of smoothness indicators are given in Appendix~\ref{sec:smoothness-indicators}.} are designed such that, in smooth regions $\beta^{(3,k)} = \mathcal{O}(\Delta x^2)$ while a substencil $\mathrm{S}^{(3,k)}$ containing discontinuities would yield $\beta^{(3,k)} = \mathcal{O}(1)$. Accordingly, the WENO5-Z scheme defines the nonlinear weights as:
\begin{equation}\label{eq:nonlinear-weights}
    w^{(k)} = \frac{\tilde{w}^{(k)}}{\tilde{w}^{(0)} + \tilde{w}^{(1)} + \tilde{w}^{(2)}}, \
    \tilde{w}^{(k)} = d^{(k)} \left[
    1 + \left(
    \frac{\tau^{(5)}}{\beta^{(3,k)}+\epsilon}
    \right)^q
    \right],
\end{equation}
where $\tau^{(5)} = |\beta^{(3,0)}-\beta^{(3,2)}|$ measures the global smoothness of stencil $\mathrm{S}^{(5)}$, the parameter $q = 2$ controls the sensitivity of the adaptive weighting strategy, and $\epsilon=10^{-6}$ is a small positive number to prevent division by zero. Consequently, these nonlinear weights converge to the linear weights $d^{(k)}$ in smooth regions, while assigning negligibly small weights to non-smooth substencils $\mathrm{S}^{(3,k)}$ with large $\beta^{(3,k)}$. This adaptive procedure allows the WENO-Z scheme to achieve robust shock capturing while preserving high-order accuracy at local extrema.

\subsection{The TENO-N Scheme}

Notwithstanding their robust shock-capturing capabilities, the WENO schemes often suffer from excessive numerical dissipation, particularly in high-wavenumber regimes \citep{Pirozzoli2006, Fu2023}. To address this, the targeted ENO (TENO) scheme, originally proposed by \citet{Fu2016}, adopts a novel ENO-like stencil selection strategy.

The TENO-N scheme \citep{Fu2018}, in particular, replaces the convex combination of candidate polynomials in WENO with a novel discontinuity-location detector. Based on the normalized smoothness indicators
\begin{equation}
    \chi^{(k)} = \frac{\gamma^{(k)}}{\gamma^{(0)} + \gamma^{(1)} + \gamma^{(2)}}, \
    \gamma^{(k)} = \left(
    1 + \frac{\tau^{(5)}}{\beta^{(3,k)}+\epsilon}
    \right)^q,
\end{equation}
where $q=6$ is adopted for strong scale separation, a binary cutoff function
\begin{equation}
    \delta^{(3, k)} = \begin{cases}
    0,& \text{if}\ \chi^{(k)} <C_{\mathrm{T}} \\
    1,& \text{otherwise}
\end{cases}
\end{equation}
is applied to detect non-smooth substencils. Here, $C_{\mathrm{T}}$ is a threshold for discontinuity, and is set to $10^{-7}$ by default. A substencil $\mathrm{S}^{(3,k)}$ is considered to contain discontinuities if $\delta^{(3, k)} =0$, and the target reconstruction polynomial is selected based on the values of $\{\delta^{(3, k)}\}$. For instance, consider the case where $\delta^{(3, 0)}_i = \delta^{(3, 1)}_i = 1$ and $\delta^{(3, 2)}_i = 0$; this implies that a discontinuity is located within the interval $(x_{i+1}, x_{i+2})$, as illustrated in Figure~\ref{fig:weno5}, and a polynomial excluding this discontinuity, e.g.,
\begin{equation}
    u_{i+1/2} = \frac{1}{12}\left(\overline{u}_{i-2} - 5\overline{u}_{i-1} + 13\overline{u}_{i} + 3\overline{u}_{i+1}\right),
\end{equation}
is adopted for the reconstruction of the cell interface value at $x_{i+1/2}$. The complete set of candidate polynomials for the TENO-N schemes can be found in \citet{Fu2018}.

\subsection{Smoothness Indicators}\label{sec:smoothness-indicators}

The smoothness indicators for the WENO and TENO schemes, originally proposed by \citet{Jiang1996}, are defined as:
\begin{equation}
\beta_i^{(r,k)} = \sum_{l=1}^{r-1}(\Delta x)^{2l-1}\int_{x_{i-1/2}}^{x_{i+1/2}}
\left[
\frac{\mathrm{d}^l}{\mathrm{d}x^l}p_i^{(r,k)}(x)
\right]^2 \mathrm{d}x,
\end{equation}
where $p^{(r,k)}_i(x)$ represents the reconstruction polynomial associated with the substencil $\mathrm{S}^{(r,k)}_i$ in a $(2r-1)$-point stencil $\mathrm{S}^{(2r-1)}_i$. In the classical WENO5 and TENO5 schemes, the explicit forms of smoothness indicators with $r=3$ are given by \citep{Jiang1996, Shu2009}:
\begin{equation}
\begin{aligned}
\beta^{(3,0)}_i &= \frac{13}{12}\left(\overline{u}_{i-2}-2\overline{u}_{i-1}+\overline{u}_i\right)^2
+\frac{1}{4}\left(\overline{u}_{i-2}-4\overline{u}_{i-1}+3\overline{u}_i\right)^2, \\
\beta^{(3,1)}_i &= \frac{13}{12}\left(\overline{u}_{i-1}-2\overline{u}_i+\overline{u}_{i+1}\right)^2
+\frac{1}{4}\left(\overline{u}_{i-1}-\overline{u}_{i+1}\right)^2, \\
\beta^{(3,2)}_i &= \frac{13}{12}\left(\overline{u}_i-2\overline{u}_{i+1}+\overline{u}_{i+2}\right)^2
+\frac{1}{4}\left(3\overline{u}_i-4\overline{u}_{i+1}+\overline{u}_{i+2}\right)^2.
\end{aligned}
\end{equation}
For the seven-point stencil multi-stencil discontinuity detector (MSDD, see Section~\ref{sec:msdd}) used in the proposed TCS7-M scheme (see Section~\ref{sec:tcs} and Appendix~\ref{appendix:tcs}), the explicit forms of smoothness indicators with $r=4$ are \citep{Balsara2000, Don2013}:
\begin{equation}
\begin{aligned}
\beta^{(4,0)}_i = \frac{1}{240}\bigl[ &
\overline{u}_{i-3}\left(547\overline{u}_{i-3}-3882\overline{u}_{i-2}+4642\overline{u}_{i-1}-1854\overline{u}_i\right)
+\overline{u}_{i-2}\left(7043\overline{u}_{i-2}-17246\overline{u}_{i-1}+7042\overline{u}_i\right) \\
+&\overline{u}_{i-1}\left(11003\overline{u}_{i-1}-9402\overline{u}_i\right)
+2107\overline{u}_i^2\bigr], \\
\beta^{(4,1)}_i = \frac{1}{240}\bigl[ &
\overline{u}_{i-2}\left(267\overline{u}_{i-2}-1642\overline{u}_{i-1}+1602\overline{u}_i-494\overline{u}_{i+1}\right)
+\overline{u}_{i-1}\left(2843\overline{u}_{i-1}-5966\overline{u}_i+1922\overline{u}_{i+1}\right) \\
+&\overline{u}_i\left(3443\overline{u}_i-2522\overline{u}_{i+1}\right)
+547\overline{u}_{i+1}^2\bigr], \\
\beta^{(4,2)}_i = \frac{1}{240}\bigl[ &
\overline{u}_{i-1}\left(547\overline{u}_{i-1}-2522\overline{u}_i+1922\overline{u}_{i+1}-494\overline{u}_{i+2}\right)
+\overline{u}_i\left(3443\overline{u}_i-5966\overline{u}_{i+1}+1602\overline{u}_{i+2}\right) \\
+&\overline{u}_{i+1}\left(2843\overline{u}_{i+1}-1642\overline{u}_{i+2}\right)
+267\overline{u}_{i+2}^2\bigr], \\
\beta^{(4,3)}_i = \frac{1}{240}\bigl[ &
\overline{u}_i\left(2107\overline{u}_i-9402\overline{u}_{i+1}+7042\overline{u}_{i+2}-1854\overline{u}_{i+3}\right)
+\overline{u}_{i+1}\left(11003\overline{u}_{i+1}-17246\overline{u}_{i+2}+4642\overline{u}_{i+3}\right) \\
+&\overline{u}_{i+2}\left(7043\overline{u}_{i+2}-3882\overline{u}_{i+3}\right)
+547\overline{u}_{i+3}^2\bigr].
\end{aligned}
\end{equation}

\section{Energy Transport Equations}\label{sec:transport}

The transport equations for kinetic and magnetic energy densities can be derived from the resistive-viscous MHD equations:
\begin{gather}
    \label{appendix:continuity} \frac{\partial \rho}{\partial t} + \nabla \cdot (\rho \bm{u}) = 0, \\
    \label{appendix:momentum} \rho \left(\frac{\partial \bm{u}}{\partial t} + \bm{u} \cdot \nabla \bm{u} \right) =
    - \nabla p
    + \bm{J} \times \bm{B} + \bm{f}
    + \nu \nabla \cdot \mathbb{T}, \\
    \label{appendix:induction} \frac{\partial \bm{B}}{\partial t} = \nabla \times (\bm{u} \times \bm{B}) + \eta \nabla^2 \bm{B}, \\
    \nabla \cdot \bm{B} = 0,
\end{gather}
where $\bm{f}$ represents all the body forces, such as the artificial driving force employed in driven MHD turbulence simulations~\eqref{eq:momentum-driven} or the Coriolis force present in shearing-box simulations~\eqref{eq:mri-momentum}. Taking the inner product of Equation~\eqref{appendix:induction} with the magnetic field vector $\bm{B}$ yields the magnetic energy transport equation (METE):
\begin{equation}\label{eq:METE}
    \frac{\partial}{\partial t}\left(\frac{1}{2} B^2\right) = \bm{B} \cdot [\nabla \times (\bm{u}\times \bm{B})] + \eta\bm{B} \cdot \nabla^2\bm{B}.
\end{equation}
Similarly, the component-wise METE can be obtained by multiplying the $i$-th component of Equation~\eqref{appendix:induction} by $B_i$:
\begin{equation}\label{eq:METE-component}
    \frac{\partial}{\partial t}\left(\frac{1}{2} B_i^2\right) = B_i[\nabla \times (\bm{u}\times \bm{B})]_i + \eta B_i \nabla^2 B_i.
\end{equation}
To avoid ambiguity, the Einstein summation convention is not adopted throughout this paper; repeated indices (e.g., $i$ in the equation above) do not imply summation. Instead, the notation $\sum$ will be explicitly used whenever a summation is intended.

To derive the kinetic energy transport equation (KETE), the continuity equation~\eqref{appendix:continuity} is first expressed in component form:
\begin{equation}
    \frac{\partial \rho}{\partial t} = - \sum_{j=1}^3 \frac{\partial (\rho u_j)}{\partial x_j}.
\end{equation}
Then, the time derivative of the $i$-direction kinetic energy density can be expressed as:
\begin{equation}\label{eq:KE-time-derivative}
    \frac{\partial}{\partial t}\left(\frac{1}{2}\rho u_i^2\right) =
    \frac{1}{2}u_i^2\frac{\partial \rho}{\partial t} + \rho u_i\frac{\partial u_i}{\partial t}
    = -\frac{1}{2}u_i^2\sum_{j=1}^3\frac{\partial(\rho u_j)}{\partial x_j} + \rho u_i\frac{\partial u_i}{\partial t}.
\end{equation}
Multiplying the $i$-th component of the momentum equation~\eqref{appendix:momentum} by $u_i$ yields:
\begin{equation}
    \rho u_i\frac{\partial u_i}{\partial t} + \rho u_i \sum_{j=1}^3 u_j \frac{\partial u_i}{\partial x_j} =
    - u_i\frac{\partial p}{\partial x_i}
    + u_i(\bm{J}\times \bm{B})_i + u_if_{i}
    + \nu u_i(\nabla \cdot \mathbb{T})_i.
\end{equation}
Substituting Equation~\eqref{eq:KE-time-derivative} into the above equation gives:
\begin{equation}
    \frac{\partial}{\partial t}\left(\frac{1}{2}\rho u_i^2\right) + \sum_{j=1}^3 \left[ \frac{1}{2}u_i^2\frac{\partial(\rho u_j)}{\partial x_j} + \rho u_i u_j \frac{\partial u_i}{\partial x_j} \right] =
    - u_i\frac{\partial p}{\partial x_i}
    + u_i(\bm{J}\times \bm{B})_i + u_if_{i}
    +\nu u_i(\nabla \cdot \mathbb{T})_i.
\end{equation}
The summation term on the left-hand side can be recast as the divergence of the $i$-th component of the kinetic energy flux:
\begin{equation}
    \nabla \cdot \left(\frac{1}{2}\rho u_i^2\bm{u}\right)
    = \sum_{j=1}^3 \left[ \frac{1}{2}u_i^2\frac{\partial(\rho u_j)}{\partial x_j}  + \rho u_iu_j\frac{\partial u_i}{\partial x_j} \right].
\end{equation}
Consequently, the component-wise KETE is given by:
\begin{equation}\label{eq:KETE-component}
    \frac{\partial}{\partial t}\left(\frac{1}{2}\rho u_i^2\right) +\nabla \cdot \left(\frac{1}{2}\rho u_i^2\bm{u}\right)  =
    - u_i\frac{\partial p}{\partial x_i}
    + u_i(\bm{J}\times \bm{B})_i + u_if_{i}
    +\nu u_i(\nabla \cdot \mathbb{T})_i,
\end{equation}
and the KETE is obtained by summing over all three components ($i=1,2,3$) of Equation~\eqref{eq:KETE-component}:
\begin{equation}
    \frac{\partial}{\partial t}\left(\frac{1}{2}\rho u^2\right) +\nabla \cdot \left(\frac{1}{2}\rho u^2\bm{u}\right)  =
    - \bm{u} \cdot \nabla p
    + \bm{u}\cdot(\bm{J}\times \bm{B}) + \bm{u} \cdot \bm{f}
    +\nu \bm{u}\cdot (\nabla \cdot \mathbb{T}).
\end{equation}

Furthermore, the modified equations for KETE and METE with numerical dissipation terms $\bm{D}^{\mathrm{num}}_{\mathrm{vis}}$ and $\bm{D}^{\mathrm{num}}_{\mathrm{res}}$ (from Equation~\eqref{eq:modified-induction-equation}, for instance) are given by:
\begin{equation}\label{eq:KETE-modified}
    \frac{\partial}{\partial t}\left(\frac{1}{2}\rho u^2\right) +\nabla \cdot \left(\frac{1}{2}\rho u^2\bm{u}\right)  =
    - \bm{u} \cdot \nabla p
    + \bm{u}\cdot(\bm{J}\times \bm{B}) + \bm{u} \cdot \bm{f}
    +\nu \bm{u}\cdot (\nabla \cdot \mathbb{T}) + \varepsilon^{\mathrm{num}}_{\mathrm{vis}},
\end{equation}
\begin{equation}\label{eq:METE-modified}
    \frac{\partial}{\partial t}\left(\frac{1}{2} B^2\right) = \bm{B} \cdot [\nabla \times (\bm{u}\times \bm{B})] + \eta\bm{B} \cdot \nabla^2\bm{B} + \varepsilon^{\mathrm{num}}_{\mathrm{res}},
\end{equation}
where
\begin{equation}
    \varepsilon^{\mathrm{num}}_{\mathrm{vis}} = \bm{u}\cdot \bm{D}^{\mathrm{num}}_{\mathrm{vis}}, \
    \varepsilon^{\mathrm{num}}_{\mathrm{res}} = \bm{B}\cdot \bm{D}^{\mathrm{num}}_{\mathrm{res}}.
\end{equation}
The corresponding component-wise equations are:
\begin{equation}\label{eq:KETE-component-modified}
    \frac{\partial}{\partial t}\left(\frac{1}{2}\rho u_i^2\right) +\nabla \cdot \left(\frac{1}{2}\rho u_i^2\bm{u}\right)  =
    - u_i\frac{\partial p}{\partial x_i}
    + u_i(\bm{J} \times \bm{B})_i + u_if_{i}
    +\nu u_i(\nabla \cdot \mathbb{T})_i + \mathscr{D}^{\mathrm{num}}_{\mathrm{vis},i},
\end{equation}
\begin{equation}\label{eq:METE-component-modified}
    \frac{\partial}{\partial t}\left(\frac{1}{2} B_i^2\right) = B_i[\nabla \times (\bm{u}\times \bm{B})]_i + \eta B_i \nabla^2 B_i + \mathscr{D}^{\mathrm{num}}_{\mathrm{res},i},
\end{equation}
where
\begin{equation}
    \mathscr{D}^{\mathrm{num}}_{\mathrm{vis},i} = u_iD^{\mathrm{num}}_{\mathrm{vis},i}, \
    \mathscr{D}^{\mathrm{num}}_{\mathrm{res},i} = B_iD^{\mathrm{num}}_{\mathrm{res},i}.
\end{equation}
For consistency, we denote the physical dissipation terms as:
\begin{equation}
    \bm{D}^{\mathrm{phy}}_{\mathrm{vis}} = \nu \nabla \cdot \mathbb{T}, \
    \bm{D}^{\mathrm{phy}}_{\mathrm{res}} = \eta \nabla^2\bm{B},
\end{equation}
yielding the physical dissipation rates
\begin{equation}
    \varepsilon^{\mathrm{phy}}_{\mathrm{vis}} = \bm{u}\cdot \bm{D}^{\mathrm{phy}}_{\mathrm{vis}}, \
    \varepsilon^{\mathrm{phy}}_{\mathrm{res}} = \bm{B}\cdot \bm{D}^{\mathrm{phy}}_{\mathrm{res}},
\end{equation}
and the component-wise physical dissipation rates
\begin{equation}
    \mathscr{D}^{\mathrm{phy}}_{\mathrm{vis},i} = u_iD^{\mathrm{phy}}_{\mathrm{vis},i}, \
    \mathscr{D}^{\mathrm{phy}}_{\mathrm{res},i} = B_iD^{\mathrm{phy}}_{\mathrm{res},i}.
\end{equation}

\section{Approximate Dispersion Relation Analysis} \label{sec:adr}

The approximate dispersion relation (ADR) analysis is widely adopted for evaluating the spectral resolution of nonlinear schemes \citep{Pirozzoli2006}. Consider a discrete harmonic function defined on a uniform grid $\{x_j=j\Delta x \mid j=0,1,\cdots, N-1\}$:
\begin{equation}
    u_j = \cos(\kappa j+\varphi),
\end{equation}
where $\kappa = 2\pi m/N$ ($m=0,1,\cdots, N/2-1$) is the normalized wavenumber, and $\varphi \in[0, 2\pi)$ is an arbitrary phase. For a general nonlinear finite difference scheme, the numerical spatial derivative can be written in the form
\begin{equation}
    \left(\frac{\partial u}{\partial x}\right)_j = \frac{1}{\Delta x}\sum_{l = -q}^{r} a_l(\{u_j\}) u_{j+l},
\end{equation}
where the coefficients $a_l$ depend (nonlinearly) on the local stencil $\{u_{j-q}, \cdots, u_{j+r}\}$.

The discrete Fourier transform (DFT) of $u_j$ into Fourier space is given by
\begin{equation}
    \tilde{u}_m = \mathrm{DFT}\{u_j\} = \frac{1}{2} \mathrm{e}^{i\varphi}\delta_{m, m_\kappa} + \frac{1}{2} \mathrm{e}^{-i\varphi}\delta_{m, -m_\kappa},
\end{equation}
where $m_\kappa = (N/2\pi)\kappa$ and $\delta_{m, n}$ is the Kronecker delta function. An ideal numerical scheme free of dispersion and dissipation errors (e.g., the spectral difference scheme) would exactly reproduce the analytical derivative in spectral space, satisfying
\begin{equation}
    \mathrm{DFT}\left\{\left(\frac{\partial u}{\partial x}\right)_j \right\} = \frac{i\kappa}{\Delta x} \tilde{u}_m.
\end{equation}
To quantify the deviation of the numerical derivative from the exact one, the modified (normalized) wavenumber $\mathrm{K}$ is defined as \citep{Fauconnier2011}:
\begin{equation}
    \mathrm{K}(\kappa) = \frac{\Delta x}{i\tilde{u}_{m_\kappa}} \mathrm{DFT}\left\{\left(\frac{\partial u}{\partial x}\right)_j \right\} \Bigg|_{m = m_\kappa} = -2i \mathrm{e}^{-i\varphi}\Delta x\, \mathrm{DFT}\left\{\left(\frac{\partial u}{\partial x}\right)_j \right\} \Bigg|_{m = m_\kappa},
\end{equation}
where the DFT of the numerical derivative is evaluated as
\begin{equation}
    \mathrm{DFT}\left\{\left(\frac{\partial u}{\partial x}\right)_j \right\} = \frac{1}{N}\sum_{j=0}^{N-1} \left(\frac{\partial u}{\partial x}\right)_j \mathrm{e}^{-i \kappa j}.
\end{equation}
Substituting this into the definition yields the explicit expression for the modified wavenumber:
\begin{equation}
    \mathrm{K}(\kappa) = -2i \mathrm{e}^{-i\varphi}\frac{\Delta x}{N}\sum_{j=0}^{N-1} \left(\frac{\partial u}{\partial x}\right)_j \mathrm{e}^{-i \kappa j}.
\end{equation}

From this relation, the numerical dispersion and dissipation properties of the scheme are characterized by the real part $\mathrm{Re}(\mathrm{K})$ and the imaginary part $\mathrm{Im}(\mathrm{K})$ as functions of $\kappa$, respectively. In principle, the resulting ADR depends on the phase $\varphi$ for nonlinear schemes. Therefore, in practice, the dispersion relation $\mathrm{K}(\kappa)$ is evaluated over a large ensemble of random phases to obtain a statistically representative result.

\section{Cell Averages to Cell Centers} \label{sec:avg2ctr6}

To derive a sixth-order transformation from cell averages to cell centers, we perform a sixth-order truncated three-dimensional Taylor series expansion of the function $u(x,y,z)$ around the cell center $(x_i, y_j, z_k)$ and integrate over the cell volume $\Omega_{i,j,k}$. This establishes a relationship between the cell averages and the cell centers along with their derivatives with sixth-order accuracy:
\begin{equation}\label{eq:ctr2avg6}
    \begin{aligned}
        \overline{u}_{i,j,k} = u_{i,j,k}
        &
        + \frac{1}{24}\left. \left[
        (\Delta x)^2 \frac{\partial^2 u}{\partial x^2}
        + (\Delta y)^2 \frac{\partial^2 u}{\partial y^2}
        + (\Delta z)^2 \frac{\partial^2 u}{\partial z^2}
        \right] \right|_{i,j,k}
        + \frac{1}{1920}\left. \left[
        (\Delta x)^4 \frac{\partial^4 u}{\partial x^4}
        + (\Delta y)^4 \frac{\partial^4 u}{\partial y^4}
        + (\Delta z)^4 \frac{\partial^4 u}{\partial z^4}
        \right] \right|_{i,j,k} \\
        &
        + \frac{1}{576}\left. \left[
        (\Delta x\Delta y)^2 \frac{\partial^4 u}{\partial x^2\partial y^2}
        + (\Delta x\Delta z)^2 \frac{\partial^4 u}{\partial x^2\partial z^2}
        + (\Delta y\Delta z)^2 \frac{\partial^4 u}{\partial y^2\partial z^2}
        \right] \right|_{i,j,k}
        + \mathcal{O}(\Delta_i^6).
    \end{aligned}
\end{equation}
To invert this relationship and express $u_{i,j,k}$ in terms of cell averages, we apply a similar fourth-order truncated expansion
\begin{equation}
\overline{(\cdot)}_{i,j,k} = (\cdot)_{i,j,k}
+ \frac{1}{24}\left. \left[
  (\Delta x)^2 \frac{\partial^2 (\cdot)}{\partial x^2}
+ (\Delta y)^2 \frac{\partial^2 (\cdot)}{\partial y^2}
+ (\Delta z)^2 \frac{\partial^2 (\cdot)}{\partial z^2}
\right] \right|_{i,j,k} + \mathcal{O}(\Delta_i^4)
\end{equation}
to the second-order partial derivatives (e.g., ${\partial^2 u}/{\partial x^2}$), and substitute them back into Equation~\eqref{eq:ctr2avg6}, yielding:
\begin{equation}
\begin{aligned}
u_{i,j,k} = \overline{u}_{i,j,k}
&
- \frac{1}{24}\left. \left[
(\Delta x)^2 \overline{
\frac{\partial^2 u}{\partial x^2} }
+ (\Delta y)^2\overline{\frac{\partial^2 u}{\partial y^2} }
+ (\Delta z)^2\overline{\frac{\partial^2 u}{\partial z^2} }
\right] \right|_{i,j,k}
+ \frac{7}{5760}\left. \left[
  (\Delta x)^4\overline{\frac{\partial^4 u}{\partial x^4} }
+ (\Delta y)^4\overline{\frac{\partial^4 u}{\partial y^4} }
+ (\Delta z)^4\overline{\frac{\partial^4 u}{\partial z^4} }
\right] \right|_{i,j,k} \\
&
+ \frac{1}{576}\left. \left[
  (\Delta x\Delta y)^2\overline{\frac{\partial^4 u}{\partial x^2\partial y^2} }
+ (\Delta x\Delta z)^2\overline{\frac{\partial^4 u}{\partial x^2\partial z^2} }
+ (\Delta y\Delta z)^2\overline{\frac{\partial^4 u}{\partial y^2\partial z^2} }
\right] \right|_{i,j,k}
+ \mathcal{O}(\Delta_i^6).
\end{aligned}
\end{equation}
By neglecting terms of order $\mathcal{O}(\Delta_i^6)$ and higher, we obtain a transformation from cell averages to cell centers with sixth-order accuracy.

\section{Targeted Compact Scheme} \label{appendix:tcs}

For smooth stencils, the TCS7-M scheme adopts the central compact scheme (Equation~\eqref{eq:central-compact-scheme})
\begin{equation}
    \beta u'_{i-2} + \alpha u'_{i-1} + u'_i + \alpha u'_{i+1} + \beta u'_{i+2} =
    c \frac{u_{i+3} - u_{i-3}}{6\Delta x} + b \frac{u_{i+2} - u_{i-2}}{4\Delta x} + a \frac{u_{i+1} - u_{i-1}}{2\Delta x}
\end{equation}
with
\begin{equation}
    \alpha, \beta = 0.5771439, 0.0896406;\
    a, b, c = 1.3025166, 0.9935500, 0.03750245
\end{equation}
taken from the spectrally optimized scheme (3.1.6) of \citet{Lele1992}. When discontinuities are detected (by MSDD) in $[u_{i-3},u_{i-2}]$ and/or $[u_{i+2},u_{i+3}]$, the eighth-order central compact scheme (2.1.12) of \citet{Lele1992} with
\begin{equation}
    \alpha, \beta = \frac{4}{9}, \frac{1}{36};\
    a, b, c = \frac{40}{27}, \frac{25}{54}, 0
\end{equation}
is employed instead. For other discontinuity configurations, the equations for TCS7-M are derived from \citet{Jiang2001} and take the general form
\begin{equation}
    \beta_- u'_{i-2} + \alpha_- u'_{i-1} + u'_i + \alpha_+ u'_{i+1} + \beta_+ u'_{i+2} = \frac{1}{\Delta x} \left(b_- u_{i-2} + a_- u_{i-1} + c_0 u_i + a_+ u_{i+1} + b_+ u_{i+2}\right),
\end{equation}
where the free parameter $\vartheta$ in Equation (2) of \citet{Jiang2001} is set to $3/10$ to achieve the highest order of accuracy. Specifically, for stencils containing a single discontinuity in $[u_{i-2},u_{i-1}]$, the coefficients are given by:
\begin{equation}
    \beta_-, \alpha_-, \alpha_+, \beta_+ = 0, \frac{11}{49}, \frac{24}{49}, \frac{3}{98}; \
    b_-, a_-, c_0, a_+, b_+ = 0, -\frac{33}{49}, \frac{53}{196}, \frac{40}{49}, \frac{25}{196}.
\end{equation}
If a single discontinuity is located in $[u_{i-1},u_i]$, the coefficients become:
\begin{equation}
    \beta_-, \alpha_-, \alpha_+, \beta_+ = 0, 0, \frac{13}{5}, \frac{3}{10}; \
    b_-, a_-, c_0, a_+, b_+ = 0, 0, -\frac{53}{20}, \frac{7}{5}, \frac{5}{4}.
\end{equation}
If a single discontinuity is located in $[u_{i},u_{i+1}]$, the coefficients become:
\begin{equation}
    \beta_-, \alpha_-, \alpha_+, \beta_+ = \frac{3}{10}, \frac{13}{5}, 0, 0; \
    b_-, a_-, c_0, a_+, b_+ = -\frac{5}{4}, -\frac{7}{5}, \frac{53}{20}, 0, 0.
\end{equation}
Finally, if a single discontinuity is located in $[u_{i+1},u_{i+2}]$, the coefficients are:
\begin{equation}
    \beta_-, \alpha_-, \alpha_+, \beta_+ = \frac{3}{98}, \frac{24}{49}, \frac{11}{49}, 0; \
    b_-, a_-, c_0, a_+, b_+ = -\frac{25}{196}, -\frac{40}{49}, -\frac{53}{196}, \frac{33}{49}, 0.
\end{equation}

Applying the equations of TCS7-M at all $N$ grid points yields a sparse linear system of $N$ equations for the unknown derivatives $u_i'$. In the present work, this system is solved using the biconjugate gradient stabilized method (BiCGSTAB, \citet{vanderVorst1992}) via \texttt{jax.scipy.sparse.linalg.bicgstab}, as implemented in our Python package \texttt{PyMHD}. The initial guess for the iterative solver, corresponding to \texttt{x0} in \texttt{bicgstab}, is taken as the approximate derivative computed using the TENO-M scheme introduced in Section~\ref{sec:msdd}, thereby accelerating convergence. Moreover, compared with the original \texttt{SciPy} implementation, the \texttt{JAX} implementation is more convenient for the direct extension to three-dimensional grids.


\bibliography{reference}{}
\bibliographystyle{aasjournalv7}



\end{CJK*}
\end{document}